\documentclass[11pt,letterpaper]{article}

\usepackage{times}
\usepackage{url}
\usepackage{graphicx}
\usepackage{subfigure}
\usepackage{color}
\usepackage{epsfig}
\usepackage{latexsym}

\usepackage{amssymb}
\usepackage{amsthm}
\usepackage{framed}
\usepackage{a4wide}
\usepackage{enumerate}
\usepackage{makeidx}
\usepackage{bm}
\usepackage{epsfig}
\usepackage{graphics}
\usepackage{colortbl}
\usepackage{colordvi}
\usepackage{verbatim}
\usepackage{amsmath}
\usepackage{multirow} 
\usepackage{booktabs} 

\newcommand{\probconverge}{\mbox{~$\stackrel{P}{\longrightarrow}$}}

\newcommand {\C} {{\rm I\kern-5.5pt C}}

\newcommand{\bP}[1]{{\mathbb{P}}\left[{#1}\right]}
\newcommand{\bE}[1]{{\mathbb{E}}\left[{#1}\right]}

\newcommand{\1}[1]{{\bf 1}\left[#1\right]}       

\newcommand{\fsquare}{\vrule height6pt width7pt depth1pt}   
\newcommand{\myproof}{{\hfill \\ \bf Proof. \ }}           
\newcommand{\myendpf}{\hfill\fsquare \\[0.1in]}             
\newcommand{\myvec}[1]{{\mbox{\boldmath{$#1$}}}}

\newtheorem{theorem}{Theorem}[section]

\newtheorem{lemma}[theorem]{Lemma}
\newtheorem{proposition}[theorem]{Proposition}

\newtheorem{fact}[theorem]{Fact}

\begin{document}

\sloppy

\title{On the log-normality of the degree distribution\\
        in large homogeneous binary \\
         multiplicative attribute graph models}

\author{Sikai Qu and Armand M. Makowski\\
        Department of Electrical and  Computer Engineering,\\
        and Institute for Systems Research \\
        University of Maryland, College Park, MD 20742.\\
        Email: skqu@umd.edu, armand@isr.umd.edu\\
}

\maketitle

\noindent
\hrulefill\\
The muliplicative attribute graph (MAG) model was introduced by Kim and Leskovec \cite{KimLeskovec}  as a mathematically tractable 
model for networks where
network structure is believed to be shaped by features or attributes associated with individual nodes.
For large homogeneous binary MAGs, they argued through approximation arguments that 
the \lq\lq tail of [the] degree distribution follows a log-normal distribution"
as the number of nodes becomes unboundedly large and the number of attributes scales logarithmically with the number of nodes.
Under the same limiting regime, we revisit the asymptotic behavior of the degree distribution:
Under weaker conditions than in \cite{KimLeskovec} 
we obtain a {\em precise} convergence result to log-normality, develop from it {\em reasoned} log-normal approximations to the degree distribution 
and derive various rates of convergence.
In particular, we show that a certain transformation of the node degree converges in distribution to a log-normal distribution,
and give its convergence rate in the form of  a Berry-Esseen type estimate. 

\noindent
\hrulefill\\

\pagebreak

\section{Introduction}
\label{sec:Introduction}

In many networks of interest, there is good reason to argue that
network structure is shaped by features or attributes associated with individual nodes.
The {\em multiplicative attribute graph} (MAG) model was recently introduced by Kim and Leskovec \cite{KimLeskovec} 
as a mathematically tractable model which implements this basic idea.
MAG models are a special case of {\em hidden variable} models discussed in earlier literature where
each node is endowed with a set of intrinsic (\lq\lq hidden") attributes, e.g., authority, social success, wealth, etc.,
and the creation of a link between two nodes expresses a mutual \lq\lq benefit" based on their attributes, e.g. see references
\cite{BogunaPastor, CCDM,SC, Sodeberg_2002, YoungSchneinerman} for examples.

In the version of the MAG model considered here, each node has $L$ {\em binary} attributes, say $0$ or $1$, organized into a $\{0,1\}^L$-valued random variable (rv).
The propensity for two nodes to have an (undirected) link between them is then determined by a probability
which  jointly (and symmetrically) depends on the attribute vectors of the two
nodes. See Section \ref{sec:MAG+ModelDefinition} for a formal construction of this model
under strong independence and homogeneity assumptions -- The resulting random graph is an instance of the {\em homogeneous binary} MAG,
and we use the notation $\mathbb{M}(n;L)$ to refer to it when the network comprises $n$ nodes and each node has $L$ binary attributes.

In \cite{KimLeskovec} Kim and Leskovec studied several aspects of this model, including its connectivity, the existence of a giant component,
the network diameter and the degree distribution. This was done in the asymptotic regime 
when the number $n$ of nodes and the number $L$ of attributes both grow unboundedly large, the latter scaling with the former as
$L_n \sim \rho \ln n$ for some $\rho > 0$ (in which case the scaling $n \rightarrow L_n$ is said to be $\rho$-admissible).

For the random graph $\mathbb{M}(n;L)$, the link variables form an exchangeable family of rvs, 
and the degree rvs of the $n$ nodes are therefore
equidistributed -- Let $D_{n,L}$ denote the generic degree of a node in $\mathbb{M}(n;L)$.
 Kim and Leskovec argue that the \lq\lq tail of [the] degree distribution follows a log-normal distribution"
when the number of nodes becomes unboundedly large and the number of attributes scales according to $\rho$-admissible scalings
\cite[Thm. 6.1, p. 129]{KimLeskovec}. Their arguments are based on the normal approximation of Binomial distributions,
and exploit the fact that the rv $D_{n,L}$ is conditionally Binomial given a Binomial rv ${\rm Bin}(L, \mu(1))$ where $\mu(1)$ is the probability that
a single attribute assumes the value $1$; see Proposition \ref{prop:DegreePMF} in Section \ref{sec:DegreeRVs+pmf}.
Log-normality of the degree distribution has been observed in a number of datasets, e.g., \cite{LeskovecHorvitz,LibenNowell2005},
and contrasts with the seemingly more prevalent power-law behavior found in many networks,
e.g., see the discussion in the monographs \cite[Section 4.2]{Durrett_Book} and \cite[Chapter 1]{vanderHofstad},
and in the survey paper \cite{NewmanSurvey}.
The fact that under certain conditions an homogeneous binary MAG generates
a network with a log-normal degree distribution points to its modeling flexibility as a generative network model.

In the present paper we further investigate the limiting properties of the degree distribution under admissible scalings.
Under weaker conditions than in \cite{KimLeskovec} 
we obtain sharper results in the form of {\em precise} convergence results, attending approximations and various rates of convergence.
The contributions can be summarized as follows: 
Consider a $\rho$-admisible scaling $\myvec{L}: \mathbb{N}_0 \rightarrow \mathbb{N}_0: n \rightarrow L_n$ for some $\rho >0$,
hence there exists a sequence $\myvec{\rho}: \mathbb{N}_0 \rightarrow \mathbb{R}_+: n \rightarrow \rho_n$ 
is determined through $L_n = \rho_n \ln n$ for each $n=1,2, \ldots $ with
$\lim_{n \rightarrow \infty} \rho_n = \rho$.

\begin{enumerate}

\item[(i)] In Lemma \ref{lem:TrivialLimits} we show that the sequence of rvs $\{ D_{n,L_n}, \ n=2,3, \ldots \}$ 
admits essentially only {\em trivial} limits;
see Section \ref{sec:ProofLimitProbDeg=0} for proofs.

This set the stage for the remainder of the paper where the condition
\begin{equation}
1 + \rho \log \Gamma(1)^{\mu(1)} \Gamma(0)^{\mu(0)} > 0
\label{eq:KeyConditionAgain}
\end{equation}
is assumed with $\mu(0) = 1 - \mu(1)$, and $\Gamma(1)$ and $\Gamma(0)$ are the quantities (\ref{eq:Gamma}) naturally associated with the MAG model.

\item[(ii)] Whereas the degree rvs themselves have only trivial limits, 
Theorem \ref{thm:LOGNORMAL} shows that  the sequence of transformed rvs
\begin{equation}
\left \{ 
(D_{n,L_n})^{\frac{1}{\sqrt{L_n}}}e^{-\sqrt{L_n}\left(\frac{1}{\rho_n} + \ln \Gamma(1)^{\mu(1)}\Gamma(0)^{\mu(0)} \right ) },
\ n=2,3, \ldots 
\right \}
\label{eq:Transformed}
\end{equation}
converges {\em in distribution} to a log-normal distribution.
Theorem  \ref{thm:LOGNORMAL}  is established in Sections \ref{sec:KeyApproximationResult} and Section \ref{sec:Proof:ThmLogNormal},
and is essentially a byproduct of an underlying Central Limit Theorem.

\item[(iii)] Theorem  \ref{thm:LOGNORMAL}  can then be exploited in Section \ref{sec:ConsequencesMainResult+LogNormal} along two different directions. 
In Section  \ref{subsec:RateConvergence}. we first derive from it a {\em rate of convergence} that complements
Lemma \ref{lem:TrivialLimits}  under condition (\ref{eq:KeyConditionAgain}), namely that the sequence of ratios 
\[
\left \{ \frac{  D_{n,L_n} }{ n^{ 1 + \rho_n \ln \Gamma(1)^{\mu(1)}\Gamma(0)^{\mu(0)} } } , \ n=2,3, \ldots \right \}
\]
has a weak limit (as an extended rv) which we identify.
In Section \ref{subsec:LogNormalApproximation}  we also extract from Theorem  \ref{thm:LOGNORMAL} a log-normal {\em approximation} to the pmf
$\{ \bP{ D_{n,L_n} =d } , \ d=0,1, \ldots \}$ when $n$ is large, namely an approximation to $\bP{ D_{n,L_n} = d }$ for each $d=1,2, \ldots $.
In Section \ref{subsec:ComparingKLvsQM} we discuss how the results obtained here relate to the original result of Kim and Leskovec.

\item[(iv)] Finally, in the same way that the Berry-Esseen estimate gives a rate of convergence for the classical Central Limit Theorem, we 
derive a Berry-Esseen type estimate for the aforementioned convergence of the sequence
(\ref{eq:Transformed}) to log-normality obtained in Theorem \ref{thm:LOGNORMAL}.
This is the content of Theorem \ref{thm:BerryEsseenLogNormal}
discussed in Section \ref{sec:BerryEsseenLogNormal}; its proof is organized in 
Section \ref{sec:DECOMPOSITION}, Section  \ref{sec:AuxiliaryResults} and Section \ref{sec:ProofTheoremBerryEsseenLogNormal}.

\end{enumerate}
Some of the results presented here were initially reported in the Ph.D. thesis of Qu \cite{Qu_PhThesis}, sometimes with different proofs.
A word on the notation and conventions in use: Unless specified otherwise,
all limiting statements, including asymptotic equivalences, are understood with $n$ going to infinity. 
The rvs under consideration are all 
defined on the same probability triple  $(\Omega, {\cal F}, \mathbb{P})$. 
The construction of a probability triple sufficiently large to carry all required rvs is standard,
and omitted in the interest of brevity.
All probabilistic statements are made with respect to the probability
measure $\mathbb{P}$, and we denote the corresponding expectation operator by $\mathbb{E}$. 
We abbreviate almost sure(ly) (under $\mathbb{P}$) by a.s.
The notation $\probconverge_n$ (resp.  $\Longrightarrow_n$) 
is used to signify convergence in probability
(resp. convergence in distribution) (under $\mathbb{P}$) with $n$ going to infinity;
see the monographs \cite{BillingsleyBook, ChungBook, ShiryayevBook} for definitions and properties.
If $E$ is a subset of $\Omega$, then $\1{E}$ is the indicator rv
of the set $E$ with the usual understanding that $\1{E}(\omega) = 1$ (resp. $\1{E}(\omega) = 0$) if $\omega \in E$
(resp. $\omega \notin E$).
The symbol $\mathbb{N}$ (resp. $\mathbb{N}_0$)  denotes the set of non-negative (resp. positive) integers.
We view sequences as mappings defined on $\mathbb{N}_0$; the mapping itself is denoted by
bolding the symbol used for the generic element of the corresponding sequence.
Unless otherwise specified, all logarithms are natural logarithms with $\ln x$ denoting the natural logarithm of $x > 0$.

\section{Homogeneous (binary) MAG models}
\label{sec:MAG+ModelDefinition}

The MAG  model is parametrized by a number of quantities, chief amongst
them the number $n$ of nodes present in the network
and the number $L$ of binary attributes associated with each node -- Both $n$ and $L$ are positive integers.
Nodes are labeled $u=1,2, \ldots$, while attributes are labeled $\ell =1,2, \ldots$.
Each of the $L$ attributes associated with
a node is assumed to be binary in nature with $1$ (resp. $0$) signifying that the attribute is present (resp. absent).
We conveniently organize these $L$ attributes into a vector element $\myvec{a}_L= (a_1, \ldots , a_L)$ of  $\{0,1\}^L$.

\subsection{The underlying random variables}
\label{subsec:BasicRVs}

The propensity of nodes to attach to each other is governed by their attributes in a way to be clarified shortly.
To formalize this notion, we follow the approach adopted by Kim and Leskovec \cite{KimLeskovec}, the
construction used here being equivalent to the one found there.

The probability triple $(\Omega, {\cal F}, \mathbb{P} )$ is assumed to carry two collections of rvs, namely the  collection
\[
\left \{ A, A_\ell, A_\ell (u), \ \ell =1,2, \ldots ; \ u=1,2, \ldots
\right \}
\]
and the triangular array
\[
\left \{ U(u,v), \ u=1,2, \ldots ; \ v =u+1, u+2, \ldots  \right \} .
\]
The following assumptions are enforced throughout:

\begin{enumerate}
\item[(i)] 
The collection
$\left \{ A, A_\ell, A_\ell (u), \ \ell =1,2, \ldots ; \ u=1,2, \ldots \right \}$
and the triangular array
$\left \{ U(u,v), \ u=1,2, \ldots ; \ v =u+1, u+2, \ldots \right \}$
are {\em mutually independent};

\item[(ii)]
The rvs
$\left \{ U(u,v), \ u=1,2, \ldots ; \ v =u+1, u+2, \ldots \right \}$
are  {\em i.i.d.} rvs, 
each of which is {\em uniformly} distributed on the open interval $(0,1)$;
and 

\item[(iii)] The rvs
$\left \{ A, A_\ell, A_\ell (u), \ \ell =1,2, \ldots ; \ u=1,2, \ldots \right \}$
form a collection of {\em i.i.d.} $\{0,1\}$-valued rvs
with pmf $\myvec{\mu} = (\mu(0), \mu(1) )$ where
$\bP{ A = 0 } = \mu (0)$ and $\bP{ A = 1 } = \mu (1)$.
To avoid trivial situations of limited interest, we assume that both $\mu (0)$ and $\mu(1)$ are elements of the
{\em open} interval $(0,1)$ such that $\mu(0) + \mu(1) =1 $.
\end{enumerate} 
For each $L=1,2, \ldots $, we write
\[
\myvec{A}_L (u) = ( A_1(u) , \ldots , A_L(u) ),
\quad u=1,2, \ldots .
\]
Under the enforced assumptions,  the $\{0,1\}^L$-valued rvs
$ \left \{ \myvec{A}_L(u) , \ u=1,2, \ldots \right \} $
are i.i.d. rvs, each with i.i.d. components distributed like the generic rv $A$.
We shall also have use for the partial sum rvs
\begin{equation}
S_L (u) = A_1(u) + \ldots + A_L(u),
\quad u=1, 2, \ldots 
\label{eq:CountAttributes}
\end{equation}
For each $\ell=1,, \ldots $, we shall say that node $u$ exhibits (resp. does not exhibit) the $\ell^{th}$ attribute if $A_\ell(u)=1$ (resp. $A_\ell(u)=0$).
In that terminology, the rv $S_L(u)$ then counts the number of attributes exhibited by node $u$ amongst the first $L$ attributes.
Under the enforced assumptions, the rvs $\{ S_L (u), \ u=1,2, \ldots \}$ form a sequence of i.i.d. rvs, 
each being distributed according to the rv $A_1+ \ldots + A_L$ which is itself a Binomial rv ${\rm Bin}(L,\mu(1))$.

For notational reasons we find it convenient to augment the triangular array of uniform rvs into the larger collection
$\left \{ U(u,v), \ u,v=1,2, \ldots \right \} $ through  the definitions
\[
U (u,u) = 1
\quad \mbox{and} \quad
U(v,u) = U(u,v),
\quad 
\begin{array}{c}
v = u+1, \ldots \\
u=1,2, \ldots
\end{array}
\]

\subsection{Adjacency }
\label{subsec:MAG-Adjacency}

On the way to defining homogeneous MAGs we introduce notions of {\em adjacency} between nodes based on their attributes.
To do so we start with  an $2 \times 2$ matrix $\mathcal{Q}$ given by
\[
\mathcal{Q} 
=
( q(a,b) )
= \left (
\begin{array}{cc}
q(1,1) & q(1,0) \\
q(0,1) & q(0,0) \\
\end{array}
\right ).
\]
Throughout we assume the symmetry condition  $q(1,0) = q(0,1)$
together with the non-degeneracy conditions
$0 < q(a,b) < 1$ for all $a,b$ in $\{0,1\}$.

Fix $L=1,2, \ldots $. With the symmetric $2 \times 2$ matrix $\mathcal{Q}$ we associate 
a mapping  $Q_L : \{0,1\}^L \times \{0,1\}^L \rightarrow [0,1]$ given by
\begin{equation}
Q_L(  \myvec{a}_L,  \myvec{b}_L )
= \prod_{\ell=1}^L q(a_\ell, b_\ell),
\quad 
\myvec{a}_L,  \myvec{b}_L \in \{0,1\}^L .
\label{eq:Q_L}
\end{equation}
Interpretations for these quantities will be given shortly.
The enforced assumptions on $\mathcal{Q}$ readily imply
\begin{equation}
Q_L(  \myvec{b}_L,  \myvec{a}_L )
=
Q_L(  \myvec{a}_L,  \myvec{b}_L )
\quad \mbox{with} \quad
0  < Q_L(  \myvec{a}_L,  \myvec{b}_L ) < 1,
\quad \myvec{a}_L,  \myvec{b}_L \in \{ 0,1\}^L.
\label{eq:InheritedConditions}
\end{equation}

Pick two nodes $u,v =1,2, \ldots $. 
We say that node $u$ is  {\em $L$-adjacent} to node $v$, written $u \sim_L v$, if the condition
\begin{equation}
U(u,v) \leq Q_L ( \myvec{A}_L(u), \myvec{A}_L(v) )
\label{eq:L-Adjacency}
\end{equation}
holds, in which case an (undirected) edge from node $u$ to node $v$ is said to exist.
Obviously, $L$-adjacency is a binary relation on the set of all nodes.
Since $U(u,v) = U(v,u)$, it is plain from the symmetry condition in (\ref{eq:InheritedConditions})
that node $u$ is $L$-adjacent to node $v$ if and only if
node $v$ is $L$-adjacent to node $u$ -- This allows us to say that nodes $u$ and $v$ are $L$-adjacent
without any risk of confusion. 
Node $u$ cannot be $L$-adjacent to itself because $U(u,u) = 1$ (by convention) and $Q_L ( \myvec{A}_L(u), \myvec{A}_L(u) ) < 1$ by
(\ref{eq:InheritedConditions}) -- In other words,  $L$-adjacency does not give rise to self-loops.

We encode $L$-adjacency  through the $\{0,1\}$-valued edge rvs
$ \left \{ \chi_L(u,v), \ u,v =1, 2, \ldots  \right \} $ given by
\begin{eqnarray}
\chi_L(u,v)
=  \1{ U(u,v) \leq Q_L ( \myvec{A}_L(u), \myvec{A}_L(v) ) },
\quad u,v =1,2, \ldots 
\label{eq:EdgeAssignmentVariables}
\end{eqnarray}
with $\chi_L(u,v)=1$ (resp. $\chi_L(u,v)=0$) 
corresponding to the existence (resp. absence) of an (undirected) edge between node $u$ and node $v$.
In view of earlier remarks,  the conditions
\begin{equation}
\chi_L(u,u) = 0
\quad \mbox{and} \quad
\chi_L(v,u) = \chi_L(u,v),
\quad u,v=1,2, \ldots
\label{eq:Chi+Conditions}
\end{equation}
are all satisfied.

\subsection{Defining the binary homogeneous MAG models}

Fix $n=1,2, \ldots$ and $L=1,2, \ldots $.
The (binary homogeneous) MAG over a set of $n$ nodes, labeled $1, \ldots, n$, with
each node having $L$ attributes, labeled $1, \ldots , L$, is the random graph
$\mathbb{M}(n;L)$ whose edge set is determined through the rvs
$\left \{ \chi_L(u,v), \ u,v=1,2, \ldots, n \right \}$. 
The edges in $\mathbb{M}(n;L)$ 
are undirected and there are no self-loops, hence any realization of $\mathbb{M}(n;L)$  is a simple graph;
this is a simple consequence of (\ref{eq:Chi+Conditions}).
We shall write $V_n = \{1, \ldots , n \}$ to denote the set of nodes in $\mathbb{M}(n;L)$.

This definition is equivalent to the one given by Kim and Leskovec \cite{KimLeskovec}.\footnote{Strictly speaking, the definition given above
is slightly more restrictive than the one proposed in \cite{KimLeskovec} as we have eliminated by {\em construction} the possibility of self-loops,
whereas such links are neglected  by Kim and Leskovec  as making no contributions in the limiting regime. 
See the discussion after Theorem 3.1 in \cite{KimLeskovec}.}
In particular, with the help of Assumptions (i) and (ii),
it is a simple matter to check from (\ref{eq:EdgeAssignmentVariables}) that  the triangular array of rvs
\[
\left \{
\chi_L(u,v), \ 
\begin{array}{c} 
u,v  \in V_n \\
u < v \\
\end{array}
\right \}
\]
are  {\em conditionally independent} $\{ 0, 1 \}$-valued rvs given the i.i.d. attribute random vectors
$\{ \myvec{A}_L(w), \  w \in V_n \}$ with
\begin{eqnarray}
\lefteqn{ 
\bP{ \chi_L(u,v) = b_{uv} , 
\begin{array}{c}
u,v \in V_n \\
u < v \\
\end{array}
\Bigg | 
\myvec{A}_L(w), \  w \in V_n }
}
& &
\nonumber \\
&=&
\prod_{u,v \in V_n , \ u < v}
\bP{ \chi_L(u,v) = b_{uv}
\Big | 
\myvec{A}_L(w), \  w \in V_n },
\quad
\begin{array}{c}
b_{uv} \in \{0,1\} \\
u < v \\
u,v \in V_n. \\
\end{array}
\label{eq:ConditionsForMAG+0}
\end{eqnarray}
Moreover, it is easy to check that
\begin{eqnarray}
\lefteqn{
\bP{ \chi_L(u,v)= 1  | \myvec{A}_L(w) ,  \  w \in V_n } 
} & &
\nonumber \\
&=&
\bP{ U(u,v) \leq Q_L ( \myvec{A}_L(u), \myvec{A}_L(v) )  | \myvec{A}_L(w) ,  \  w \in V_n }
\nonumber \\
&=& 
Q_L(  \myvec{A}_L(u),  \myvec{A}_L(v) ),
\quad
\begin{array}{c}
u \neq v \\
u,v \in V_n \\
\end{array}
\label{eq:ConditionsForMAG}
\end{eqnarray}
with the symmetric mapping 
$Q_L : \{0,1\}^L \times \{0,1\}^L \rightarrow [0,1]$ being the one introduced earlier at (\ref{eq:Q_L}).
The probabilistic characteristics of $\mathbb{M}(n,L)$ are 
completely determined by the matrix $\mathcal{Q}$ and by the pmf $\myvec{\mu}$. 
These building blocks are assumed given and held {\em fixed} during the discussion -- They will not
be explicitly displayed in the notation.

\section{Degree rvs in homogeneous MAGs and their common pmf}
\label{sec:DegreeRVs+pmf}

Fix $n=2,3, \ldots $ and $L=1,2, \ldots $.
For each $u$ in $V_n$, the degree rv $D_{n,L}(u)$ of node $u$ in $\mathbb{M}(n,L)$ is the rv given by
\begin{equation}
D_{n,L}(u)= \sum_{v \in V_n, \ v\neq u}\chi_L(u,v).
\label{eq:DegreeRV1}
\end{equation}
It counts the number of one-hop neighbors of node $u$ in $\mathbb{M}(n;L)$, and can be expressed as
\begin{equation}
D_{n,L}(u)= \sum_{v \in V_n, \ v\neq u} \1{ U(u,v) \leq Q_L ( \myvec{A}_L(u), \myvec{A}_L(v) ) }
\label{eq:DegreeRV2}
\end{equation}
with the help of (\ref{eq:EdgeAssignmentVariables}).

We begin by identifying the probability mass function (pmf) of the rvs defined at (\ref{eq:DegreeRV2}).
We shall find it convenient to express this result and others with the help of the quantities $\Gamma(0)$  and $\Gamma(1) $ defined by
\begin{equation}
\Gamma(0) =  \bE{ q(0, A) } 
\quad \mbox{and} \quad
\Gamma(1) = \bE{ q(1, A) } .
\label{eq:Gamma}
\end{equation}

\begin{proposition}
{\sl
Fix $n=2,3,\dots$ and $L=1,2,\dots$. For each $u$ in $V_n$, the degree rv $D_{n,L}(u)$ is distributed according to
a compound binomial distribution: The conditional distribution of the rv $D_{n,L}(u)$ given $S_L(u)$ is that of
a Binomial rv  ${\rm Bin}( n-1, \Gamma(1)^{S_{L}(u)}\Gamma(0)^{L-S_L(u)})$, namely
\begin{eqnarray}
\lefteqn{
\bP{D_{n,L}(u)=d | S_{L}(u)=\ell } 
} & &
\nonumber \\
&=&
 \binom{n-1}{d} \left(\Gamma(1)^{\ell}\Gamma(0)^{L-\ell}\right)^{d}\left(1-\Gamma(1)^{\ell}\Gamma(0)^{L-\ell}\right)^{n-1-d},
 \quad
 \begin{array}{c}
 d=0, 1, \ldots , n-1 \\
 \ell =0,1, \ldots , L \\
 \end{array}
\label{eq:DegreeConditionalPMF}
\end{eqnarray}
with  
\begin{equation}
\bP{ S_L(u) = \ell } =  \binom{L}{\ell} \mu(1)^\ell \mu(0)^{L - \ell },
\quad  \ell =0,1, \ldots , L.
\label{eq:Binomial}
\end{equation}
}
\label{prop:DegreePMF}
\end{proposition}

This result was given by Kim and Leskovec \cite[Cor. 10.8, p. 151]{KimLeskovec} as a corollary of a more general 
result of Young and  Schneinerman \cite[Section 5, p. 142]{YoungSchneinerman}. Here we provide a direct proof in the notation of the paper.

\myproof
Fix $n=2,3,\dots$ and $L=1,2,\dots$, and pick $u$ in $V_n$.
In what follows, $b_v$ is an arbitrary element in $\{0,1\}$ for each $v$ in $V_n$.
With the help of Assumptions (i) and (ii) the arguments that justify
(\ref{eq:ConditionsForMAG+0}) and (\ref{eq:ConditionsForMAG}) also give
\begin{eqnarray}
\lefteqn{ 
\bP{ \chi_L(u,v) = b_v, 
\begin{array}{c}
v \in V_n \\
v \neq u \\
\end{array}
\Bigg | 
\myvec{A}_L(w), \  w \in V_n }
}
& &
\nonumber \\
&=&
\prod_{ v \in V_n , v \neq u } 
\left (
b_v Q_L( \myvec{A}_L(u), \myvec{A}_L(v))  + (1-b_v) \left (  1 - Q_L( \myvec{A}_L(u), \myvec{A}_L(v)) \right )
\right ) 
\end{eqnarray}
and that 
\begin{eqnarray}
\lefteqn{
\bP{ \chi_L(u,v) = b_v \Bigg |  \myvec{A}_L(w), \  w \in V_n }  } & &
\nonumber \\
&=&
b_v Q_L( \myvec{A}_L(u), \myvec{A}_L(v))  + (1-b_v) \left (  1 - Q_L( \myvec{A}_L(u), \myvec{A}_L(v)) \right ),
\quad v \in V_n .
\nonumber
\end{eqnarray}

Next, iterated conditioning and standard conditioning arguments yield the a.s. equalities
\begin{eqnarray}
& &
\bP{ \chi_L (u,v) = b_v , 
\begin{array}{c}
v \neq u \\
v \in V_n \\
\end{array}
\Big | \myvec{A}_L(u)  } 
\nonumber \\
&=&
\bE{
\prod_{v \in V_n, v \neq u} 
\left (
b_v Q_L ( \myvec{A}_L(u),  \myvec{A}_L(v) ) 
+ (1-b_v) ( 1 - Q_L ( \myvec{A}_L(u),  \myvec{A}_L(v) ) )
\right )
\Bigg | \myvec{A}_L(u)  } 
\nonumber \\
&=&
\bE{
\prod_{v \in V_n, v \neq u} 
\left (
b_v Q_L ( \myvec{a}_L,  \myvec{A}_L(v) ) 
+ (1-b_v) ( 1 - Q_L ( \myvec{a}_L,  \myvec{A}_L(v) ) )
\right )
\Bigg | \myvec{A}_L(u)  }_{\myvec{a}_L =  \myvec{A}_L(u)}
\nonumber \\
&=&
\bE{
\prod_{v \in V_n, v \neq u} 
\left (
b_v Q_L ( \myvec{a}_L,  \myvec{A}_L(v) ) 
+ (1-b_v) ( 1 - Q_L ( \myvec{a}_L,  \myvec{A}_L(v) ) )
\right )
}_{\myvec{a}_L =  \myvec{A}_L(u)}
\nonumber \\
&=&
\left ( \prod_{v \in V_n, v \neq u} 
\bE{
b_v Q_L ( \myvec{a}_L,  \myvec{A}_L(v) ) 
+ (1-b_v) ( 1 - Q_L ( \myvec{a}_L,  \myvec{A}_L(v) ) )
}
\right )_{\myvec{a}_L =  \myvec{A}_L(u)}
\nonumber \\
&=&
\prod_{v \in V_n, v \neq u} 
\left (
b_v \bE{ Q_L ( \myvec{a}_L,  \myvec{A}_L(v) ) }_{\myvec{a}_L =  \myvec{A}_L(u)}
+ (1-b_v) ( 1 - \bE{ Q_L ( \myvec{a}_L,  \myvec{A}_L(v) ) }_{\myvec{a}_L =  \myvec{A}_L(u)}
\right )
\nonumber
\end{eqnarray}
where we made repeated use of the fact that the rvs
$\{ \myvec{A}_L(w), \ w \in V_n \}$ are i.i.d. rvs.

For each $v$ in $V_n$, the components $A_1(v), \ldots , A_L(v)$ being i.i.d. rvs, we conclude that
\begin{eqnarray}
 \bE{ Q_L ( \myvec{a}_L ,  \myvec{A}_L(v) ) }
&=&
 \bE{ \prod_{\ell=1}^L q(a_\ell, A_\ell(v) ) }
\nonumber \\
&=&
\prod_{\ell=1}^L \bE{ q(a_\ell, A_\ell (v) ) }
\nonumber \\
&=&
\bE{ q(1, A) }^{\sum_{\ell=1}^L a_\ell} \cdot  \bE{ q( 0, A) }^{\sum_{\ell=1}^L (1-a_\ell)},
\quad \myvec{a}_L \in \{0,1\}^L 
\nonumber
\end{eqnarray}
with the $\{0,1\}$-valued rv $A$ denoting a generic representative  of the i.i.d. rvs $A_1(v), \ldots , A_L(v)$. 
The relation 
\[
 \bE{ Q_L ( \myvec{a}_L ,  \myvec{A}_L(v) ) }_{  \myvec{a}_L = \myvec{A}_L(u) } 
 =
 \Gamma(1)^{S_L(u)} \Gamma(0)^{L-S_L(u)}
\]
immediately follows, and reporting this fact into earlier expressions yields the a.s equality
\begin{eqnarray}
& &
\bP{ \chi_L (u,v) = b_v , 
\begin{array}{c}
v \neq u \\
v \in V_n \\
\end{array}
\Big | \myvec{A}_L(u)  } 
\nonumber \\
&=&
\prod_{v \in V_n, v \neq u} 
\left (
b_v \Gamma(1)^{S_L(u)} \Gamma(0)^{L-S_L(u)} + (1-b_v)  \left ( 1 - \Gamma(1)^{S_L(u)} \Gamma(0)^{L-S_L(u)} \right )
\right ).
\nonumber
\end{eqnarray}
Therefore, the $\{0,1\}$-valued rvs
$\{ \chi_L(u,v), \ v \neq u , v \in V_n \}$ are conditionally i.i.d. given $\myvec{A}_L(u)$ with
\[
\bP{ \chi_L (u,v) = 1 \Big | \myvec{A}_L(u)  } 
= \Gamma(1)^{S_L(u)} \Gamma(0)^{L-S_L(u)}
\quad \mbox{a.s.}
\]
The rv $S_L(u)$ being $\sigma ( \myvec{A}_L(u)  )$-measurable, the $\{0,1\}$-valued rvs
$\{ \chi_L(u,v), \ v \neq u , v \in V_n \}$ are also conditionally i.i.d. given $S_L(u)$ with
\begin{equation}
\bP{ \chi_L (u,v) = 1 \Big | S_L(u)  } 
= \Gamma(1)^{S_L(u)} \Gamma(0)^{L-S_L(u)}
\quad \mbox{a.s.}
\end{equation}
The conclusion of Proposition \ref{prop:DegreePMF} immediately follows.

\myendpf

Specializing Proposition \ref{prop:DegreePMF} for $d=0$ yields the expression
\[
\bP{D_{n,L}(u) = 0 | S_{L}(u) = \ell } 
=
\left(1-\Gamma(1)^{\ell}\Gamma(0)^{L-\ell}\right)^{n-1},
\quad  \ell =0,1, \ldots , L 
\]
whence
\begin{equation}
\bP{D_{n,L}(u) = 0 }
= 
\bE{ \left(1-\Gamma(1)^{S_L(u) }\Gamma(0)^{L-S_L(u)}\right)^{n-1} }
\label{eq:DegreePMFwith:d=0}
\end{equation}
upon taking expectations.

Under the enforced assumptions and the symmetry condition in (\ref{eq:InheritedConditions}), 
it is plain from the expressions (\ref{eq:DegreeRV2}) that the rvs
$ D_{n,L}(1), \ldots , D_{n,L}(n)$ are exchangeable rvs,
hence the probability distribution of the rv $ D_{n,L}(u)$ does not depend on $u$ in $V_n$.
Most of the results will be given in terms of the  {\em generic} rv $D_{n,L}$ with the understanding that
it is a rv with the same distribution as that of any of the rvs $ D_{n,L}(1), \ldots , D_{n,L}(n)$ -- We shall refer to $D_{n,L}$ 
as  {\em the} degree rv in $\mathbb{M}(n;L)$.
However, for sake of concreteness, the arguments  in the proofs  will use $D_{n,L}(1)$ as a convenient representative for $D_{n,L}$; we will do so
without any further mention to this probabilistic equivalence.

\section{Main results I: Log-normality}
\label{sec:MainResult+LogNormal}

The following terminology will simplify the presentation of the results:
A {\em scaling} (for the number of attributes) is any mapping $\myvec{L}: \mathbb{N}_0 \rightarrow \mathbb{N}_0: n \rightarrow L_n$.
With $\rho >0$, the scaling $\myvec{L}: \mathbb{N}_0 \rightarrow \mathbb{N}_0$ is said to be $\rho$-{\em admissible} if 
\begin{equation}
L_n \sim \rho \ln  n,
\label{eq:Rho-Admissible}
\end{equation}
in which case it holds that
\begin{equation}
L_n = \rho_n \ln  n,
\quad n=1,2, \ldots 
\label{eq:Rho-AdmissibleEquivalent}
\end{equation}
for some sequence  $\myvec{\rho}: \mathbb{N}_0 \rightarrow \mathbb{R}_+: n \rightarrow \rho_n$ 
such that $\lim_{n \rightarrow \infty} \rho_n = \rho$. The sequence $\myvec{\rho}: \mathbb{N}_0 \rightarrow \mathbb{R}_+$
defined by (\ref{eq:Rho-AdmissibleEquivalent}) is uniquely determined by the $\rho$-scaling $\myvec{L}: \mathbb{N}_0 \rightarrow \mathbb{N}_0$,
and is said to be {\em associated} with it.

Interest in admissible scalings is discussed in \cite{KimLeskovec}.
The definition of admissibility given by Kim and Leskovec \cite{KimLeskovec} uses logarithms in base two;
results given there are easily reconciled with the results presented here through the well-known fact
that $\ln x = \ln 2 \cdot \log_2 x $ where $\log_2 x $ is the logarithm of $x > 0$ in base $2$.

Isolated nodes play a key role in the forthcoming discussion.
This is captured by the following technical fact 

\begin{lemma}
{\sl With $\rho > 0$, for any $\rho-$admissible scaling $\myvec{L}:\mathbb{N}_0 \rightarrow\mathbb{N}_0$,
the zero-one law
\begin{equation}
\lim_{n \rightarrow \infty} \bP{D_{n,L_n} = 0 }
= 
\left \{
\begin{array}{ll}
1 & \mbox{if $~1+ \rho \ln \Gamma(1)^{\mu(1)} \Gamma(0)^{\mu(0)} < 0$ } \\
   & \\
0 & \mbox{if $~1+ \rho \ln \Gamma(1)^{\mu(1)} \Gamma(0)^{\mu(0)} > 0$ } \\
\end{array}
\right .
\label{eq:LimitProbDeg=0}
\end{equation}
holds.
}
\label{lem:LimitProbDeg=0}
\end{lemma}

The proof of Lemma \ref{lem:LimitProbDeg=0} is given in Section \ref{sec:ProofLimitProbDeg=0}
where related arguments also lead to a complete characterization of the limiting degree distribution.

\begin{lemma}
{\sl 
Under any $\rho-$admissible scaling $\myvec{L}:\mathbb{N}_0 \rightarrow\mathbb{N}_0$ with $\rho > 0$,
the limiting distribution of the rvs $\{ D_{n,L_n} , \ n=2,3, \ldots \}$ is trivial in the following sense:
If $1+ \rho \ln \Gamma(1)^{\mu(1)} \Gamma(0)^{\mu(0)} < 0$, then $ D_{n,L_n} \Longrightarrow_n 0$, while
if  $1+ \rho \ln \Gamma(1)^{\mu(1)} \Gamma(0)^{\mu(0)} > 0$, then $ D_{n,L_n} \Longrightarrow_n \infty$ in that
$\lim_{n \rightarrow \infty}  \bP{ D_{n,L_n}  = d } = 0$ for each $d=0,1, \ldots $.
 }
\label{lem:TrivialLimits}
\end{lemma}

The boundary case
\[
1+ \rho \ln \Gamma(1)^{\mu(1)} \Gamma(0)^{\mu(0)} = 0
\]
will not be pursued further in this paper.
By Lemma \ref{lem:TrivialLimits}, in all other cases, the degree rvs $\{ D_{n,L_n} , \ n=2,3, \ldots \}$ themselves converge only
to a {\em trivial}  limit. This prompts us to seek limiting results that involve a transformation of these rvs.
The next result, established in Section \ref{sec:Proof:ThmLogNormal}, provides such a result, but first some notation:

Let $Z$ denote a standard Gaussian rv, a fact written $Z \sim {\rm N}(0,1)$, namely
\[
\bP{ Z \leq z } = \Phi(z)
\quad \mbox{with} \quad \Phi (z) = \int_{-\infty}^z \frac{1}{\sqrt{2\pi}} e^{- \frac{t^2}{2}} dt,
\quad z \in \mathbb{R}.
\]
With $m$ and $\sigma$ arbitrary in $\mathbb{R}$,
the log-normal distribution with parameters $(m, \sigma^2)$ is denoted by $ \ln {\rm N} \left ( m, \sigma^2 \right )$, and is the
probability distribution of the rv 
$e^{ m + \sigma Z }$.
The dependence on $\sigma$ is only through $\sigma^2$ 
as can be seen from the fact that the rvs $e^{ m + \sigma Z } $ and $e^{ m - \sigma Z } $ 
are equidistributed since the symmetric rvs $\sigma Z $ and $- \sigma Z $ have the same distribution.
It is well known that
\begin{equation}
\bP{ e^{m + \sigma Z  } \leq x }
=
\bP{ Z \leq \frac{ \ln x - m }{| \sigma |} },
\quad x \geq 0
\label{eq:LOG+CDF}
\end{equation}
with probability density function given by
\begin{equation}
\frac{d}{dx} \bP{ e^{m + \sigma Z  } \leq x } 
= 
\frac{1}{x \cdot \sqrt{ 2\pi \sigma^2 } } e^{- \frac{ \left ( \ln x - m \right )^2 } {2 \sigma^2 } },
\quad x > 0.
\label{eq:LOG+pdf}
\end{equation}

Throughout the paper the quantity $\sigma$ is given by
\begin{equation}
\sigma \equiv \sqrt{ \mu(0) \mu(1) } \cdot  \ln \frac{\Gamma(1)}{\Gamma(0)}.
\label{eq:SIGMA}
\end{equation}
Note that $\sigma > 0$ (resp. $\sigma < 0$) if $\Gamma (0) < \Gamma(1)$ (resp. $\Gamma(1) < \Gamma(0)$).

\begin{theorem}
{\sl
Assume that the condition 
\begin{equation}
1 + \rho \ln \Gamma(1)^{\mu(1)}  \Gamma(0)^{\mu(0)} > 0
\label{eq:KeyCondition}
\end{equation}
holds  for some $\rho > 0$.
For any $\rho-$admissible scaling $\myvec{L}:\mathbb{N}_0 \rightarrow\mathbb{N}_0$ with associated sequence
$\myvec{\rho}: \mathbb{N}_0 \rightarrow \mathbb{R}_+: n \rightarrow \rho_n$, we have
\begin{equation}
(D_{n,L_n})^{\frac{1}{\sqrt{L_n}}}e^{-\sqrt{L_n}\left(\frac{1}{\rho_n} + \ln \Gamma(1)^{\mu(1)}\Gamma(0)^{\mu(0)} \right ) }
\Longrightarrow_n  
\ln \rm{N}  \left(0, \sigma^2  \right) .
\label{eq:LOGNORMAL}
\end{equation}
}
\label{thm:LOGNORMAL}
\end{theorem}

The definition of weak convergence in terms of probability distribution functions allows a restating of Theorem \ref{thm:LOGNORMAL} as
\begin{equation}
\lim_{n \rightarrow \infty}
\bP{ (D_{n,L_n})^{\frac{1}{\sqrt{L_n}}}e^{-\sqrt{L_n}\left(\frac{1}{\rho_n} + \ln \Gamma(1)^{\mu(1)}\Gamma(0)^{\mu(0)} \right ) } \leq x }
=
\bP{ e^{\sigma Z } \leq x },
\quad x > 0
\label{eq:LOGNORMAL+1}
\end{equation}
since the log-normal distribution $\ln {\rm N} \left(0, \sigma^2  \right) $ has no point of discontinuity -- The range $x < 0$ is omitted 
as both probabilities vanish on that range.
Furthermore, this last convergence  occurs uniformly \cite[Thm. 1.11, p. 17]{PetrovBook} \cite{ShiryayevBook} in the sense that
\begin{equation}
\lim_{n \rightarrow \infty}
\sup_{x \geq 0 }
\left |
\bP{ (D_{n,L_n})^{\frac{1}{\sqrt{L_n}}}e^{-\sqrt{L_n}\left(\frac{1}{\rho_n} + \ln \Gamma(1)^{\mu(1)}\Gamma(0)^{\mu(0)} \right ) } \leq x  }
-
\bP{ e^{ \sigma Z } \leq x }
\right | = 0.
\label{eq:LOGNORMAL+2}
\end{equation}

\section{Consequences of Theorem \ref{thm:LOGNORMAL} }
\label{sec:ConsequencesMainResult+LogNormal}

The observation (\ref{eq:LOGNORMAL+2}) can be exploited in a number of ways as we now discuss;
the notation remains as in Section \ref{sec:MainResult+LogNormal}.
First we note that
\begin{eqnarray}
L_n \left(\frac{1}{\rho_n} + \ln \Gamma(1)^{\mu(1)}\Gamma(0)^{\mu(0)} \right )
=
\left(1 + \rho_n \ln \Gamma(1)^{\mu(1)}\Gamma(0)^{\mu(0)} \right ) \cdot \ln n ,
\quad n=2,3, \ldots
\nonumber
\end{eqnarray}
and (\ref{eq:LOGNORMAL+2}) can be restated as
\begin{equation}
\lim_{n \rightarrow \infty}
\sup_{x \geq 0 }
\left |
\bP{ \left ( 
\frac{  D_{n,L_n} }{ n^{ 1 + \rho_n \ln \Gamma(1)^{\mu(1)}\Gamma(0)^{\mu(0)} }}  \right )^\frac{1}{\sqrt{L_n}} \leq x  }
-
\bP{ e^{ \sigma Z } \leq x }
\right | = 0.
\label{eq:LOGNORMAL+3}
\end{equation}

\subsection{Complementing Lemma \ref{lem:TrivialLimits}}
\label{subsec:RateConvergence}

The mapping $\mathbb{R}_+ \rightarrow \mathbb{R}_+: x \rightarrow x^{\sqrt{L_n}}$ being a bijection, we get
\begin{equation}
\lim_{n \rightarrow \infty}
\sup_{ t \geq 0 }
\left |
\bP{
\frac{  D_{n,L_n} }{ n^{ 1 + \rho_n \ln \Gamma(1)^{\mu(1)}\Gamma(0)^{\mu(0)} }} 
 \leq  t
} -
\bP{ e^{ \sigma Z } \leq t^\frac{1}{\sqrt{L_n}} }
\right |
= 0.
\nonumber
\end{equation}
For each $t > 0$, we note that $\lim_{n \rightarrow \infty} t^\frac{1}{\sqrt{L_n}} = 1 $, whence
$\lim_{n \rightarrow \infty} \bP{ e^{ \sigma Z } \leq t^\frac{1}{\sqrt{L_n}} } = \bP{ e^{ \sigma Z }  \leq 1 }  = \bP{ \sigma Z \leq 0 } = \bP{ Z \leq 0 } = \frac{1}{2}$
regardless of the sign of $\sigma$, so that
\[
\lim_{n \rightarrow \infty} 
\bP{ \frac{  D_{n,L_n} }{ n^{ 1 + \rho_n \ln \Gamma(1)^{\mu(1)}\Gamma(0)^{\mu(0)} }}  \leq  t } 
= \frac{1}{2},
\quad t > 0 .
\]
It is easy to check that
\begin{equation}
\lim_{n \rightarrow \infty} \bP{  \frac{  D_{n,L_n} }{ n^{ 1 + \rho_n \ln \Gamma(1)^{\mu(1)}\Gamma(0)^{\mu(0)} }}  \leq  t } 
= \left \{
\begin{array}{ll}
0 & \mbox{if $t \leq 0$} \\
    &   \\
\frac{1}{2} & \mbox{if $t > 0$} \\
\end{array}
\right .
\label{eq:LOGNORMAL+9}
\end{equation}
with the case $t=0$ handled by Lemma \ref{lem:LimitProbDeg=0}.
Thus, if $\Lambda$ is an extended-valued rv that takes the values $0$ and $\infty$ with equal probability, namely
$\bP{ \Lambda = 0 } = \bP{ \Lambda = \infty} = \frac{1}{2}$,
then (\ref{eq:LOGNORMAL+9}) can be more compactly restated as
\begin{equation}
\frac{  D_{n,L_n} }{ n^{ 1 + \rho_n \ln \Gamma(1)^{\mu(1)}\Gamma(0)^{\mu(0)} } } 
\Longrightarrow_n \Lambda
\label{eq:ComplementTrivialLimits}
\end{equation}
where $\Longrightarrow_n$ is now understood as weak convergence of extended-valued rvs.
Comparing this last fact with the convergence $D_{n,L_n} \Longrightarrow_n \infty$ under the condition (\ref{eq:KeyCondition}), we 
can interpretate (\ref{eq:ComplementTrivialLimits}) as a rate of convergence complementing Lemma \ref{lem:TrivialLimits}. Interestingly,
although $\lim_{n \rightarrow \infty} n^{1 + \rho_n \ln \Gamma(1)^{\mu(1)}\Gamma(0)^{\mu(0)} } = \infty$ under (\ref{eq:KeyCondition}),
$D_{n,L_n}$ goes to infinity faster than $n^{ 1 + \rho_n \ln \Gamma(1)^{\mu(1)}\Gamma(0)^{\mu(0)} }$ with probability half,
but slower than $n^{ 1 + \rho_n \ln \Gamma(1)^{\mu(1)}\Gamma(0)^{\mu(0)} }$ with probability half.

\subsection{A log-normal approximation}
\label{subsec:LogNormalApproximation}

We can also write (\ref{eq:LOGNORMAL+3}) in the following equivalent form
\begin{equation}
\lim_{n \rightarrow \infty}
\sup_{x \geq 0 }
\left |
\bP{ D_{n,L_n} \leq x^{\sqrt{L_n}} \cdot n^{ 1 + \rho_n \ln \Gamma(1)^{\mu(1)}\Gamma(0)^{\mu(0)} }   }
-
\bP{ e^{ \sigma Z } \leq x }
\right | = 0.
\label{eq:LOGNORMAL+4}
\end{equation}
Now fix $t > 0$ and $n=2,3, \ldots $. The equation
\[
x^{\sqrt{L_n}} \cdot n^{ 1 + \rho_n \ln \Gamma(1)^{\mu(1)}\Gamma(0)^{\mu(0)} } = t,
\quad x > 0
\]
has a unique solution $x_n(t) > 0$ given by
\begin{equation}
x_n (t) 
\equiv 
\left ( t \cdot n^{ -( 1 + \rho_n \ln \Gamma(1)^{\mu(1)}\Gamma(0)^{\mu(0)} ) } \right )^{ \frac{1}{ \sqrt{L_n} }}
\label{eq:x_n(t)}
\end{equation}
(with the convention $x_n(0)=0$).
The mapping $\mathbb{R}_+ \rightarrow \mathbb{R}_+: t \rightarrow x_n(t)$ being a bijection, we can translate 
(\ref{eq:LOGNORMAL+4}) into
\begin{equation}
\lim_{n \rightarrow \infty}
\sup_{ t \geq 0 } \left |
\bP{ D_{n,L_n} \leq t }
-
\bP{ e^{\sigma Z } \leq x_n(t) }
\right |
= 0.
\label{eq:LOGNORMAL+5}
\end{equation}

The approximation
\begin{eqnarray}
\bP{ D_{n,L_n} \leq t }
=_{\rm Approx} \bP{ e^{\sigma Z } \leq x_n(t) }
= \Phi \left ( \frac{ \ln x_n(t) }{|\sigma|} \right ),
\quad t \geq  0
\label{eq:Approximation1}
\end{eqnarray}
naturally suggests itself where $=_{\rm Approx}$ means that the lefthand side is approximated by the
righthand side.
The approximation is uniform in $t > 0$, and becomes increasingly accurate as $n$ becomes large.

In particular, fix $d=1,2, \ldots $. Since
\[
\bP{ D_{n,L_n} =d  } = \bP{ D_{n,L_n} \leq d} - \bP{ D_{n,L_n} \leq d-1 },
\quad n=2,3, \ldots 
\]
we also get from (\ref{eq:LOGNORMAL+5}) that 
\[
\lim_{n \rightarrow \infty}
\sup_{d=0,1, \ldots } 
\left | \bP{ D_{n,L_n} = d }
-
\bP{ x_n(d-1) < e^{\sigma Z } \leq x_n(d) }
\right |
= 0,
\]
whence
\begin{eqnarray}
 \bP{ D_{n,L_n} = d } 
&=_{\rm Approx}&
 \bP{ x_n(d-1) < e^{\sigma Z } \leq x_n(d) }
\nonumber \\
&=&
\Phi \left ( \frac{ \ln x_n(d) }{|\sigma|} \right ) -  \Phi \left (  \frac{ \ln x_n(d-1) }{|\sigma|}\right ),
\label{eq:Approxbased CLT}
\end{eqnarray}
the approximation being uniform in $d$.
Using (\ref{eq:x_n(t)}) we note that 
\[
\ln x_n (d) 
=
\frac{1}{\sqrt{L_n}}  
\left ( \ln d -  \left(1 + \rho_n \ln \Gamma(1)^{\mu(1)}\Gamma(0)^{\mu(0)} \right ) \cdot \ln n  \right ) 
\]

\subsection{Comparison with the approximation by Kim and Leskovec }
\label{subsec:ComparingKLvsQM}

We close by comparing the results discussed here with the approximation obtained by Kim and Leskovec \cite[Thm. 6.1, p. 129]{KimLeskovec}.
In that reference, for any $\rho$-admissible scaling $\myvec{L}: \mathbb{N}_0 \rightarrow \mathbb{N}_0$ 
satisfying (\ref{eq:KeyCondition}) for some $\rho > 0$, the authors state that \lq\lq the tail of the degree distribution follows a log-normal
distribution $\ln {\rm N} ( m_{n,{\rm KL}}, \sigma^2_{n,{\rm KL} })$ as $n$ goes to infinity."
In the notation used in the present paper, for each $n=2,3, \ldots $, the parameters $m_{n,{\rm KL}}$ and $\sigma^2_{n,{\rm KL}}$ are given by
\begin{equation}
m_{n,{\rm KL} }
=
\ln \left ( n \Gamma(1)^{L_n} \right ) + L_n \cdot \mu(0) \ln R_{\rm KL}  + \frac{L_n}{2} \cdot \mu(0) \mu(1) \left ( \ln R_{\rm KL}  \right )^2
\label{eq:Parameter1}
\end{equation}
and
\begin{equation}
\sigma^2_{n,{\rm KL} }
= L_n \cdot \mu(1)\mu(0) \left ( \ln R_{\rm KL}  \right )^2 
\label{eq:Parameter2}
\end{equation}
where $ R_{\rm KL}  = \frac{ \Gamma (0) }{\Gamma(1)}$ -- The parameter $\mu$ appearing in \cite{KimLeskovec} coincides with  $\mu(0)$ (so $1-\mu = \mu(1)$).

The analysis carried out by Kim and Leskovec is done under the following assumptions, namely 
(i) the entries of the symmetric matrix $\mathcal{Q}$ are ordered as $ q(1,1) < q(0,1) = q(1,0)  < q(0,0)$; and 
(ii)  the ratio $\frac{\Gamma(0)}{\Gamma(1)} $ lies in the interval $(1.6, 3)$. 
These additional conditions were {\em not} needed to derive the results presented
here. A careful inspection of the proof \cite[Section 10.4]{KimLeskovec} shows that the authors derive an asymptotic expression (for $n$ large) for 
the quantity $\ln \bP{ D_{n,L_n} = d_n }$ when $d_n$ scales with $n$ so that $\lim_{n \rightarrow \infty}d_n = \infty$, yet $d_n = o(n)$.
This allows Stirling's approximation to be used to approximate the binomial coefficients 
${n-1 \choose d_n }$ as $n$, $d_n$ and $n-1-d_n$ all go to infinity -- See the expression (\ref{eq:DegreeConditionalPMF}).
Under theses conditions the discussion culminates in the approximation
\begin{eqnarray}
\ln \bP{ D_{n,L_n} = d_n }
\approx
C^\prime - \frac{1}{2 \sigma^2_{n,{\rm KL} } }
\left (
\ln d_n - m_{n,KL} 
\right )^2
\label{eq:LogProbability}
\end{eqnarray}
for large $n$ where  $C^\prime$ is a  positive constant, thereby forming the basis for the conclusion of 
Theorem 6.1 \cite[p. 129]{KimLeskovec}.
However, It is unclear from (\ref{eq:LogProbability}) 
whether $\ln \bP{ D_{n,L_n} = d_n }$ is approximated for large $n$  by the logarithm of 
the probability distribution $\ln {\rm N} ( m_{n,{\rm KL}}, \sigma^2_{n,{\rm KL} })$ (at $d_n$)
or by its probability density function (at $d_n$).

This results given here are significantly different from the one obtained by Kim and Leskovec in \cite{KimLeskovec} 
in that we provide
(i) a {\em precise} convergence result to log-normality in Theorem \ref{thm:LOGNORMAL}, and (ii) 
a {\em reasoned} approximation, namely (\ref{eq:Approxbased CLT}),  to the probability $\bP{ D_{n,L_n} = d }$ for {\em each} $d = 1,2, \ldots $.
In an attempt reconcile the result of Kim and Leskovec  note the following:
Starting with the expressions  (\ref{eq:Parameter1}) and (\ref{eq:Parameter2}), for all $n=1,2, \ldots $,  easy calculations  show that
\begin{eqnarray}
\sigma^2_{n,{\rm KL} }
= L_n \cdot \mu(1)\mu(0) \left (-  \ln  \frac{\Gamma(1)}{\Gamma(0)}  \right )^2 = \rho_n \sigma^2 \cdot \ln n
\end{eqnarray}
with $\sigma$ given by (\ref{eq:SIGMA}) and
\begin{eqnarray}
m_{n,{\rm KL} }
&=&
\ln \left ( n e^{ L_n \ln \Gamma(1) } \right ) - L_n \mu(0) \ln \frac{\Gamma(1)}{\Gamma(0)}   + \frac{L_n}{2} \mu(0) \mu(1) \left ( - \ln \frac{\Gamma(1)}{\Gamma(0)}  \right )^2
\nonumber \\
&=&
\ln n +  \left ( \ln \Gamma(1) -  \mu(0) \ln \frac{\Gamma(1)}{\Gamma(0)}  \right )   \cdot L_n + \frac{\sigma^2 }{2} \cdot L_n
\nonumber \\
&=& 
\left ( 1 + \rho_n \ln \Gamma(1)^{\mu(1)} \Gamma(0) ^{\mu(0)} \right ) \cdot \ln n + \frac{\sigma^2 }{2} \cdot \rho_n \ln n .
\end{eqnarray}
Next, the basic convergence  (\ref{eq:LOGNORMAL}) suggests that for large $n$, the distribution of the rv $D_{n,L_n}$ is instead
well approximated by that of the rv
\begin{eqnarray}
e^{L_n \left(\frac{1}{\rho_n} + \ln \Gamma(1)^{\mu(1)}\Gamma(0)^{\mu(0)} \right ) }  \cdot \left ( e^{\sigma Z}  \right )^{\sqrt{L_n}}
&=&
e^{ \left ( 1  + \rho_n \ln \Gamma(1)^{\mu(1)}\Gamma(0)^{\mu(0)} \right ) \ln n }  \cdot e^{\sigma \sqrt{L_n} \cdot Z} 
\nonumber \\
&=&
e^{ m_{n,{\rm KL}} - \frac{\sigma^2 }{2} \cdot \rho_n \ln n + \sigma_{n,{\rm KL}} \cdot Z} 
\nonumber \\
&=&
n^{ - \frac{\sigma^2 }{2} \rho_n }  \cdot e^{ m_{n,{\rm KL}} +  \sigma_{n,{\rm KL}}  \cdot Z} .
\end{eqnarray}

\section{Main results II: A Berry-Esseen type result}
\label{sec:BerryEsseenLogNormal}

In order to obtain a Berry-Esseen type estimate to complement Theorem \ref{thm:LOGNORMAL}, we derive bounds to the quantities
\[
\sup_{ x \geq 0 } | \Delta_n (x) |, 
\quad n=2,3, \ldots 
\]
where for each $x \geq 0 $ we have set
\begin{eqnarray}
\Delta_n (x)
\equiv 
\bP{ (D_{n,L_n})^{\frac{1}{\sqrt{L_n}}}e^{-\sqrt{L_n}\left(\frac{1}{\rho_n} + \ln \Gamma(1)^{\mu(1)}\Gamma(0)^{\mu(0)} \right ) } \leq x }
 -
\bP{   \left ( \frac{ \Gamma(1)}{\Gamma(0)}  \right )^{\sigma_ 0 Z} \leq x } 
\label{eq:DELTA}
\end{eqnarray}
and $\sigma_0$ is given by 
\begin{equation}
\sigma_0 = \sqrt{ \mu(0) \mu(1) } .
\label{eq:Constants}
\end{equation}

We shall have use for  the monotone increasing mapping $\Psi: \mathbb{R}_+ \rightarrow \mathbb{R}_+$ given by
\begin{equation}
\Psi(x) 
= (x+1) \ln (x+1) - x ,
\quad x \geq 0.
\label{eq:psi}
\end{equation}
Elementary calculus shows that
$\Psi^\prime (x) = \ln (x+1)$ and that $\Psi^{\prime\prime} (x) = (x+1)^{-1}$, whence $\Psi (0) = 0$ and $\Psi^\prime (0) = 0$, and 
\[
\Psi (x) \geq \frac{x^2}{2(1+x)},
\quad x > 0
\]
by a standard Taylor series approximation.

\begin{theorem}
{\sl
Assume  that $\Gamma(0) \neq \Gamma(1)$ and that condition (\ref{eq:KeyCondition}) holds  for some $\rho > 0$.
For any $\rho-$admissible scaling $\myvec{L}:\mathbb{N}_0 \rightarrow\mathbb{N}_0$, there exists a universal constant $C^\star$ such that
for each $n=2,3, \ldots $ we have
\begin{eqnarray}
\sup_{x \geq 0 } | \Delta_n(x) | 
&\leq &
\frac{1}{\sqrt{2\pi \sigma^2 L_n}}  \cdot \ln \left ( \frac{1+\delta}{1-\delta} \cdot \frac{n}{n-1} \right )
+
\frac{3C^\star}{\sqrt{L_n}} \cdot \frac{ \mu(1)^2 + \mu(0)^2 }{ \sqrt{\mu(1)\mu(0)}}
\nonumber \\
& &
~ + 4 e^{ -2 L_n \eta^2 }
+
2 e^{ - \Psi (\delta) \cdot (n-1) \left ( \Gamma(1)^{\mu(1) + \eta }  \Gamma(0)^{\mu(0) + \eta} \right )^{L_n} }
\label{eq:BerryEsseenLogNormal}
\end{eqnarray}
with $\delta$ arbitrary in $(0,1)$ and $\eta$ arbitrary in $(0, \mu(1))$.
}
\label{thm:BerryEsseenLogNormal}
\end{theorem} 

This result is established in Section \ref{sec:ProofTheoremBerryEsseenLogNormal} with preliminary material
discussed in Section \ref{sec:DECOMPOSITION} and Section \ref{sec:AuxiliaryResults}.
To get a better sense of the rate of convergence at (\ref{eq:BerryEsseenLogNormal}) as a function of $n$, recall that the $\rho$-admissible
scaling $\myvec{L}:\mathbb{N}_0 \rightarrow\mathbb{N}_0$ has associated sequence 
$\myvec{\rho}: \mathbb{N}_0 \rightarrow \mathbb{R}_+: n \rightarrow \rho_n$, i.e.,
$L_n =\rho_n \ln n$ for all $n=2,3, \ldots $ with $\lim_{n \rightarrow \infty} \rho_n = \rho$.
The first two terms together behave like $O ( \frac{1}{ \sqrt{\ln n}} )$. For the remaining terms,
 the usual arguments yield
 \[
e^{-2  L_n \eta^2 } = e^{-2 \rho_n \eta^2 \ln n } = n^{-2 \rho_n \eta^2 } = n^{-2 \eta^2 \rho (1+ o(1)) } = o \left ( \frac{1}{ \sqrt{\ln n}} \right )
\]
while
\[
e^{ - \Psi (\delta) \cdot (n-1) \left ( \Gamma(1)^{\mu(1) + \eta }  \Gamma(0)^{\mu(0) + \eta} \right )^{L_n} } 
\leq
e^{ - \frac{\delta^2}{2(1+\delta)} \cdot \frac{n-1}{n} n^{1 +  \rho_n \ln  \Gamma(1)^{\mu(1)+\eta} \Gamma(0)^{\mu(0) + \eta} } }
\]
with
\[
1 +  \rho_n \ln  \Gamma(1)^{\mu(1)+\eta} \Gamma(0)^{\mu(0) + \eta} 
=
1 +  \rho_n \ln  \Gamma(1)^{\mu(1)} \Gamma(0)^{\mu(0)} + \rho_n \eta \cdot \ln \Gamma(1) \Gamma(0).
\]
But $\lim_{ n \rightarrow \infty}  \left ( 1 +  \rho_n \ln  \Gamma(1)^{\mu(1)} \Gamma(0)^{\mu(0)} \right )
= 1 +  \rho \ln  \Gamma(1)^{\mu(1)} \Gamma(0)^{\mu(0)} > 0$ under condition (\ref{eq:KeyCondition}). 
Therefore, $\eta $ in $(0, \mu(1))$ can be selected small enough so that there exists some positive integer $n^\star (\eta)$ and
a positive constant $C(\rho; \eta) > 0$ such that
\[
\inf_{ n \geq  n^\star (\eta)  } \left ( 1 +  \rho_n \ln  \Gamma(1)^{\mu(1)+\eta} \Gamma(0)^{\mu(0) + \eta}  \right ) > C(\rho; \eta)  > 0,
\]
in which case it is plain that
\[
e^{ - \frac{\delta^2}{2(1+\delta)} \cdot \frac{n-1}{n} n^{1 +  \rho_n \ln  \Gamma(1)^{\mu(1)+\eta} \Gamma(0)^{\mu(0) + \eta} } }
= O \left ( e^{ - \frac{\delta^2}{4(1+\delta)}  \cdot n^{ C(\rho;\eta) } }  \right ) .
\]
We conclude that
\[
e^{ - \Psi (\delta) \cdot (n-1) \left ( \Gamma(1)^{\mu(1) + \eta }  \Gamma(0)^{\mu(0) + \eta} \right )^{L_n} } 
\leq O \left ( e^{ - \frac{\delta^2}{4(1+\delta)}  \cdot n^{ C(\rho;\eta) } }  \right )  = o \left ( \frac{1}{ \sqrt{\ln n}} \right ),
\]
Collecting all this information we conclude from (\ref{eq:BerryEsseenLogNormal}) that
\[
\sup_{x \geq 0 } | \Delta_n(x) |  \leq O \left ( \frac{1}{ \sqrt{\ln n}} \right ).
\]
Needless to say this bound on the rate of convergence is rather weak.

\section{Proofs of Lemma  \ref{lem:LimitProbDeg=0} and Lemma \ref{lem:TrivialLimits}}
\label{sec:ProofLimitProbDeg=0}

We begin with an easy technical fact  to be used on a number of occasions.

\begin{fact}
{\sl
With $\rho > 0$, for any $\rho-$admissible scaling $\myvec{L}:\mathbb{N}_0 \rightarrow\mathbb{N}_0$,
the zero-infinity law
\begin{equation}
\lim_{n \rightarrow \infty} (n-1) \Gamma(1)^{S_{L_n}(1) }\Gamma(0)^{L_n -S_{L_n}(1)  }
= 
\left \{
\begin{array}{ll}
0 & \mbox{if $~1+ \rho \ln \Gamma(1)^{\mu(1)} \Gamma(0)^{\mu(0)} < 0$ } \\
   & \\
\infty & \mbox{if $~1+ \rho \ln \Gamma(1)^{\mu(1)} \Gamma(0)^{\mu(0)} > 0$ } \\
\end{array}
\right .
\label{eq:BASIC}
\end{equation}
holds a.s.
}
\label{fact:BASIC}
\end{fact}

\myproof
With $\rho > 0$, consider a $\rho-$admissible scaling $\myvec{L}:\mathbb{N}_0 \rightarrow\mathbb{N}_0$.
For each $n=2,3, \ldots $, we have
\begin{eqnarray}
(n-1) \Gamma(1)^{S_{L_n}(1) }\Gamma(0)^{L_n -S_{L_n}(1)  }
&=&
\frac{ n-1 }{n} \cdot n  \left ( \Gamma(1)^{\frac{ S_{L_n}(1) }{L_n } }\Gamma(0)^{1 - \frac{ S_{L_n}(1)}{L_n} } \right )^{\rho_n \ln n}
\nonumber \\
&=&
\frac{ n-1 }{n} \cdot n^{ 1 + \rho_n \ln \left (\Gamma(1)^{\frac{ S_{L_n}(1) }{L_n } }\Gamma(0)^{1 - \frac{ S_{L_n}(1)}{L_n} } \right ) }
\label{eq:BASIC2}
\end{eqnarray}
where the sequence $\myvec{\rho}: \mathbb{N}_0 \rightarrow \mathbb{R}_+$ is
the unique sequence associated with the $\rho$-admissible scaling $\myvec{L}: \mathbb{N}_0 \rightarrow \mathbb{N}_0$.

The rv $S_{L_n}(1)$ being a sum of $L_n$ i.i.d. Bernoulli rvs with parameter $\mu(1)$ defined in (\ref{eq:CountAttributes}),
we conclude
\begin{equation}
\lim_{n \rightarrow \infty}  \frac{ S_{L_n} (1) }{L_n } = \mu (1) \quad {\rm a.s.}
\label{eq:SLLNs}
\end{equation}
by the Strong Law of Large Numbers. Let $n$ go to infinity in (\ref{eq:BASIC2}) on the a.s. event where (\ref{eq:SLLNs}) takes place.
We readily get (\ref{eq:BASIC}) upon observing that
\[
\lim_{n \rightarrow \infty} 
\left ( 1 + \rho_n \ln \left (\Gamma(1)^{\frac{ S_{L_n}(1) }{L_n } }\Gamma(0)^{1 - \frac{ S_{L_n}(1)}{L_n} } \right )  \right )
= 1 + \rho \ln \left ( \Gamma(1)^{\mu(1)} \Gamma(0)^{\mu(0)} \right )
 \quad {\rm a.s.}
\]
by virtue of (\ref{eq:SLLNs}).
\myendpf

\subsection{A proof of Lemma  \ref{lem:LimitProbDeg=0}}

Our point of departure is the expression (\ref{eq:DegreePMFwith:d=0}) for $u=1$.
With $\rho > 0$, consider a $\rho-$admissible scaling $\myvec{L}:\mathbb{N}_0 \rightarrow\mathbb{N}_0$.
For each $n=2,3, \ldots $, we have
\begin{equation}
\bP{D_{n,L_n}(1) = 0 }
= 
\bE{ \left(1-\Gamma(1)^{S_{L_n}(1) }\Gamma(0)^{L_n -S_{L_n} (1)}\right)^{n-1} }.
\label{eq:Expression}
\end{equation}

Next recall the following well-known fact \cite[Prop. 3.1.1, p. 116]{EKM} : For any sequence
${\bf a}: \mathbb{N}_0 \rightarrow \mathbb{R}_+$, we have
$\lim_{n \rightarrow \infty} \left ( 1 - a_n \right  )^{n} = e^{-c}$ 
for some $c$ in $[0,\infty]$ if and only if
$\lim_{n \rightarrow \infty} n a_n  = c$.
With this equivalence in mind, the a.s. convergence
\begin{equation}
\lim_{n \rightarrow \infty}  \left(1-\Gamma(1)^{S_{L_n}(1) }\Gamma(0)^{L_n -S_{L_n} (1)}\right)^{n-1} 
= \left \{
\begin{array}{ll}
1 & \mbox{if $1+ \rho \ln \Gamma(1)^{\mu(1)} \Gamma(0)^{\mu(0)} < 0$ } \\
   & \\
0 & \mbox{if $1+ \rho \ln \Gamma(1)^{\mu(1)} \Gamma(0)^{\mu(0)} > 0$ } \\
\end{array}
\right .
\label{eq:AuxiliaryLimit2}
\end{equation}
is seen to hold by Fact \ref{fact:BASIC}.
To complete the proof of Lemma  \ref{lem:LimitProbDeg=0}, let $n$ go to infinity in (\ref{eq:Expression}). We conclude to (\ref{eq:LimitProbDeg=0})
with the help of  (\ref{eq:AuxiliaryLimit2}), the interchange of limit and integration being justified by the Bounded Convergence Theorem
as we note the bounds
\[
\left |  \left(1-\Gamma(1)^{S_{L_n}(1) }\Gamma(0)^{L_n -S_{L_n} (1)}\right)^{n-1}  \right | \leq 1 ,
\quad n=2,3,  \ldots 
\]
\myendpf

\subsection{A proof of Lemma \ref{lem:TrivialLimits}}

If $1+ \rho \ln \Gamma(1)^{\mu(1)} \Gamma(0)^{\mu(0)} < 0$, then the desired convergence $ D_{n,L_n} \Longrightarrow_n 0$
is just a rephrasing of  Lemma  \ref{lem:LimitProbDeg=0}.

Now assume that  $1+ \rho \ln \Gamma(1)^{\mu(1)} \Gamma(0)^{\mu(0)} > 0$.
We shall show that $\lim_{n \rightarrow \infty} \bP{ D_{n,L_n} = d } = 0$ for each $d=0,1, \ldots $.
By the zero-part of Lemma  \ref{lem:LimitProbDeg=0} this already holds for $d=0$. Next, fix $d = 1, 2, \ldots $ and consider
$n=2,3, \ldots $ such that $n > d$. Using (\ref{eq:DegreeConditionalPMF})-(\ref{eq:Binomial}) we have
\begin{eqnarray}
\lefteqn{
\bP{D_{n,L_n}(1)=d | S_{L_n}(1) } 
} & &
\nonumber \\
&=&
\binom{n-1}{d} 
\left(\Gamma(1)^{S_{L_n}(1) }\Gamma(0)^{L_n- S_{L_n} (1) }\right)^{d}\left(1-\Gamma(1)^{ S_{L_n}(1) }\Gamma(0)^{L_n- S_{L_n}(1)}\right)^{n-1-d}
\nonumber \\
&\leq&
\binom{n-1}{d} 
\left(1-\Gamma(1)^{ S_{L_n}(1) }\Gamma(0)^{L_n- S_{L_n}(1)}\right)^{n-1-d} 
\nonumber \\
&\leq&
\frac{n^d}{d!}
\cdot e^{ - (n-1-d) \Gamma(1)^{ S_{L_n}(1) }\Gamma(0)^{L_n- S_{L_n}(1)} 
}
\nonumber \\
&=&
\frac{1}{d!} \cdot e^{ d \ln n - (n-1-d) \Gamma(1)^{ S_{L_n}(1) }\Gamma(0)^{L_n- S_{L_n}(1)} 
}.
\label{eq:Inequality}
\end{eqnarray}

By Fact \ref{fact:BASIC}, under the condition $1+ \rho \ln \Gamma(1)^{\mu(1)} \Gamma(0)^{\mu(0)} > 0$ we conclude that
\begin{equation}
\lim_{n \rightarrow \infty} 
\left ( d \ln n - (n-1-d) \Gamma(1)^{ S_{L_n}(1) }\Gamma(0)^{L_n- S_{L_n}(1)} \right )
= - \infty
\quad {\rm a.s.}
\label{eq:BASIC5}
\end{equation}
as we observe from (\ref{eq:BASIC2}) that $\lim_{n \rightarrow \infty} (n-1-d) \Gamma(1)^{ S_{L_n}(1) }\Gamma(0)^{L_n- S_{L_n}(1)} = \infty $ a.s.
polynomially fast.

Let $n$ go to infinity in (\ref{eq:Inequality}). It follows from (\ref{eq:BASIC5}) that
$\lim_{n \rightarrow \infty} \bP{ D_{n,L_n} = d | S_{L_n}(1) }  = 0$ a.s., whence $\lim_{n \rightarrow \infty} \bP{ D_{n,L_n} = d }  = 0$
by the Bounded Convergence Theorem since
\[
\bP{ D_{n,L_n} = d } = \bE{  \bP{ D_{n,L_n} = d | S_{L_n}(1) }  } ,
\quad n=2,3, \ldots
\]
\myendpf

\section{A key approximation result}
\label{sec:KeyApproximationResult}

The following technical fact will be used in the proof of Theorem \ref{thm:LOGNORMAL}
(given in Section \ref{sec:Proof:ThmLogNormal}).

\begin{lemma}
{\sl
Assume that the condition (\ref{eq:KeyCondition}) holds for some $\rho > 0$.
For any $\rho-$admissible scaling $\myvec{L}:\mathbb{N}_0 \rightarrow\mathbb{N}_0$, it holds that
\begin{equation}
\frac{D_{n,L_n}(1) }{\bE{D_{n,L_n}(1) |S_{L_n}(1)}}  \probconverge_n ~1.
\label{eq:ConvergenceRatio}
\end{equation}
\label{lem:ConvergenceRatio}
}
\end{lemma}

\myproof
Fix $n=2,3, \ldots $ and $L=1, 2, \ldots $.
By Proposition \ref{prop:DegreePMF} the conditional distribution of the rv $D_{n,L}(1)$ given $S_L(1)$ is that of
a Binomial rv  ${\rm Bin}( n-1, \Gamma(1)^{S_{L}(1)}\Gamma(0)^{L-S_L(1)})$, hence
\begin{equation}
\bE{ D_{n,L}(1) | S_L(1) } 
= (n-1) \Gamma(1)^{S_{L}(1)}\Gamma(0)^{L-S_L(1)} > 0.
\label{eq:ConditionalExpectation}
\end{equation}
Furthermore,  applying standard Chernoff-Hoeffding bounds for Binomial rvs
\cite[Section 2.1] {JansonLuczakRucinski} we conclude for each $\delta > 0$ that
\[
\bP{ \Bigg | D_{n,L}(1) - \bE{ D_{n,L}(1) | S_L(1) }  \Bigg |  ~> \delta  \cdot  \bE{ D_{n,L}(1) | S_L(1) }  \Bigg | S_L(1) }
\leq
2 e^{ - \Psi(\delta) \cdot \bE{ D_{n,L}(1)  | S_L(1) }  }
\]
with the mapping $\Psi : \mathbb{R}_+ \rightarrow \mathbb{R}_+  $ given by (\ref{eq:psi}).
Taking expectations gives
\[
\bP{ ~ \Bigg | D_{n,L}(1) - \bE{ D_{n,L}(1) | S_L(1) }  \Bigg |  ~ > \delta  \cdot  \bE{ D_{n,L}(1) | S_L(1) } ~ }
\leq
2 \bE{ e^{ - \Psi(\delta) \cdot \bE{ D_{n,L}(1) | S_L(1) }  } },
\]
or equivalently,
\begin{eqnarray}
\bP{ \Bigg | \frac{ D_{n,L}(1) }{ \bE{ D_{n,L}(1) | S_L(1) }  } - 1 \Bigg |  ~> \delta }
\leq 
2 \bE{ e^{ - \Psi (\delta) \cdot \bE{ D_{n,L}(1) | S_L(1) }  } }.
\label{eq:InequalityChernoff+1}
\end{eqnarray}

Considering this last bound along a $\rho$-admissible scaling $\myvec{L}: \mathbb{N}_0 \rightarrow \mathbb{N}_0$ gives
\begin{eqnarray}
\bP{ \Bigg | \frac{ D_{n,L_n}(1) }{ \bE{ D_{n,L_n}(1) | S_{L_n} (1) }  } - 1 \Bigg |  ~> \delta }
\leq 
2 \bE{ e^{ - \Psi(\delta) \cdot \bE{ D_{n,L_n}(1) | S_{L_n}(1) }  } },
\quad n=1,2, \ldots 
\label{eq:InequalityChernoff+2}
\end{eqnarray}
Let $n$ go to infinity in (\ref{eq:InequalityChernoff+2}).
By Fact \ref{fact:BASIC}, we have $\lim_{ n \rightarrow \infty } \bE{ D_{n,L_n}(1) | S_{L_n}(1) }  = \infty $ a.s.
under the condition (\ref{eq:KeyCondition}). It follows that
\[
\lim_{ n \rightarrow \infty } \bP{ \Bigg | \frac{ D_{n,L_n}(1) }{ \bE{ D_{n,L_n}(1) | S_{L_n} (1) }  } - 1 \Bigg |  ~> \delta } 
= 0
\]
with  the interchange of limit and integration validated the Bounded Convergence Theorem.
This establishes (\ref{eq:ConvergenceRatio}).
\myendpf

A little more can be gleaned from the proof of Lemma \ref{lem:ConvergenceRatio}.
This will be used later on in the proof of the Berry-Essen estimate associated with the
convergence of Theorem \ref{thm:LOGNORMAL}.

\begin{lemma}
{\sl Fix $n=2,3, \ldots $ and $L=1,2, \ldots $.
For each $\eta$ in $(0, \mu(1) )$ and every $\delta > 0$, it holds that
\begin{equation}
\bP{ \Bigg | \frac{ D_{n,L}(1) }{ \bE{ D_{n,L}(1) | S_L(1) }  } - 1 \Bigg |  ~> \delta }
\leq
4 e^{ -2 L \eta^2 }
+
2 e^{ - \Psi (\delta) \cdot (n-1) \left ( \Gamma(1)^{\mu(1) + \eta }  \Gamma(0)^{1 - \mu(1) + \eta} \right )^L }.
\label{eq:ConvergenceRatio+Rate}
\end{equation}
}
\label{lem:ConvergenceRatio+Rate}
\end{lemma}

\myproof
Fix $n=2,3, \ldots $ and $L=1,2, \ldots $. Pick $\eta > 0$ such that $0 < \eta < \mu(1)$.
On the event $[ | \frac{S_L(1)}{L}  - \mu(1)| \leq \eta ]$, we have the inequalities $\frac{S_L(1)}{L}  \leq \mu(1) + \eta $ and
$1 -  \frac{S_L(1)}{L} \leq  1 - \mu(1) + \eta$, whence
\[
\Gamma(1)^{S_L(1)}\Gamma(0)^{L-S_L(1)} 
 =
 \left ( \Gamma(1)^\frac{S_L(1)}{L}  \Gamma(0)^{ 1 - \frac{S_L(1)}{L}  } \right )^L
\geq \left ( \Gamma(1)^{\mu(1) + \eta }  \Gamma(0)^{1 - \mu(1) + \eta} \right )^L .
\]
It follows from (\ref{eq:ConditionalExpectation}) that
\[
\bE{ D_{n,L}(1) | S_L(1) } 
\geq  (n-1) \left ( \Gamma(1)^{\mu(1) + \eta }  \Gamma(0)^{1 - \mu(1) + \eta} \right )^L,
\]
whence
\begin{eqnarray}
\lefteqn{
\bE{ \1{ \Big | \frac{S_L(1)}{L}  - \mu(1) \Big | \leq \eta } \cdot  e^{ - \Psi (\delta) \cdot \bE{ D_{n,L}(1) | S_{L}(1) }  }  }
} & &
\nonumber \\
&\leq&
e^{ - \Psi (\delta) \cdot (n-1) \left ( \Gamma(1)^{\mu(1) + \eta }  \Gamma(0)^{1 - \mu(1) + \eta} \right )^L }
\cdot 
\bP{ \Big | \frac{S_L(1)}{L}  - \mu(1) \Big | \leq \eta }.
\label{eq:ZZZ+1}
\end{eqnarray}
On the other hand, the classical Hoeffding bound (for Binomial rvs) readily yields
\[
\bP{ \Big | \frac{S_L(1)}{L}  - \mu(1) \Big | >  \eta }
\leq 2 e^{ -2 L \eta^2 },
\]
so that
\begin{eqnarray}
\bE{ \1{ \Big | \frac{S_L(1)}{L}  - \mu(1) \Big | > \eta } \cdot  e^{ - \Psi (\delta) \cdot \bE{ D_{n,L}(1) | S_{L}(1) }  }  }
\leq \bP{ \Big | \frac{S_L(1)}{L}  - \mu(1) \Big | > \eta } \leq 2 e^{ -2 L \eta^2 } .
\label{eq:ZZZ+2}
\end{eqnarray}
Combining the inequalities (\ref{eq:ZZZ+1}) and (\ref{eq:ZZZ+2}) readily yields (\ref{eq:ConvergenceRatio+Rate})
with the help of (\ref{eq:InequalityChernoff+2}).
\myendpf

\section{A proof of Theorem \ref{thm:LOGNORMAL}}
\label{sec:Proof:ThmLogNormal}

Consider a $\rho$-admissible scaling  $\myvec{L}: \mathbb{N}_0 \rightarrow \mathbb{N}_0$
and let the sequence $\myvec{\rho}: \mathbb{N}_0 \rightarrow \mathbb{R}_+$ be
the unique sequence associated with the $\rho$-admissible scaling $\myvec{L}: \mathbb{N}_0 \rightarrow \mathbb{N}_0$,
so that  $L_n = \rho_n \ln n $ for all $n=1,2, \ldots $ with $\lim_{n \rightarrow \infty} \rho_n = \rho$.

Fix $n=2,3, \ldots $. using (\ref{eq:ConditionalExpectation}) we readily see that
\begin{eqnarray}
& &
D_{n,L_n}(1)  e^{-L_n \left(\frac{1}{\rho_n} + \ln \Gamma(1)^{\mu(1)}\Gamma(0)^{\mu(0)} \right ) } 
\nonumber \\
&=&
\left ( \frac{ D_{n,L_n}(1)}{ \bE{ D_{n,L_n} (1) | S_{L_n}(1) } }  \right )
\cdot  \bE{ D_{n,L_n} (1) | S_{L_n}(1) }
\cdot e^{-L_n \left(\frac{1}{\rho_n} + \ln \Gamma(1)^{\mu(1)}\Gamma(0)^{\mu(0)} \right ) } 
\nonumber \\
&=&
\left ( \frac{ D_{n,L_n}(1)}{ \bE{ D_{n,L_n} (1) | S_{L_n}(1) } }  \right )
\cdot \left ( (n-1) \Gamma(1)^{ S_{L_n}(1) }  \Gamma(0)^{L_n - S_{L_n}(1) }  \right )
\cdot e^{-L_n \left ( \frac{1}{\rho_n} + \ln \Gamma(1)^{\mu(1)}\Gamma(0)^{\mu(0)} \right ) } 
\nonumber \\
&=&
\frac{n-1}{n} \cdot \left ( \frac{ D_{n,L_n}(1)}{ \bE{ D_{n,L_n} (1) | S_{L_n}(1) } }  \right )
\cdot  \Gamma(1)^{ S_{L_n}(1) }  \Gamma(0)^{L_n - S_{L_n}(1) } 
\cdot \Gamma(1)^{-L_n \mu(1)}\Gamma(0)^{-L_n \mu(0)}
\nonumber \\
&=&
\frac{n-1}{n} \cdot \left ( \frac{ D_{n,L_n}(1)}{ \bE{ D_{n,L_n} (1) | S_{L_n}(1) } }  \right )
\cdot
\left ( \frac{ \Gamma(1)}{\Gamma(0)}  \right )^{ S_{L_n}(1)  - L_n \mu(1) } .
\label{eq:RELATION+1}
\end{eqnarray}
It immediately follows that
\begin{eqnarray}
\lefteqn{
(D_{n,L_n}(1) )^{\frac{1}{\sqrt{L_n}}} e^{-\sqrt{L_n}\left(\frac{1}{\rho_n} + \ln \Gamma(1)^{\mu(1)}\Gamma(0)^{\mu(0)} \right ) } 
} & &
\nonumber \\
&=&
\left ( \frac{n-1}{n} \right )^{\frac{1}{\sqrt{L_n}}}
\cdot \left ( \frac{ D_{n,L_n}(1)}{ \bE{ D_{n,L_n} (1) | S_{L_n}(1) } }  \right )^{\frac{1}{\sqrt{L_n}}}
\cdot
\left ( \frac{ \Gamma(1)}{\Gamma(0)}  \right )^{  \sqrt{L_n}  \left ( \frac{S_{L_n}(1)}{L_n}  - \mu(1) \right )} .
\nonumber
\end{eqnarray}

Let $n$ go to infinity in this last expression: It is plain that
\begin{equation}
\lim_{n \rightarrow \infty} 
\left ( \frac{n-1}{n} \right )^{\frac{1}{\sqrt{L_n}}} = 1 
\label{eq:LimitA}
\end{equation}
while under condition (\ref{eq:KeyCondition})  we have
\begin{equation}
 \left ( \frac{ D_{n,L_n}(1)}{ \bE{ D_{n,L_n} (1) | S_{L_n}(1) } }  \right )^{\frac{1}{\sqrt{L_n}}}  \probconverge_n ~1
 \label{eq:ConvergenceRatio+B}
\end{equation}
by Lemma \ref{lem:ConvergenceRatio}.
To complete the proof we note the following:
The rv $S_{L_n}(1)$ being a sum of $L_n$ i.i.d. Bernoulli rvs with parameter $\mu(1)$ 
defined in (\ref{eq:CountAttributes}) (hence with variance $\sigma_0^2$), the Central Limit Theorem yields
\begin{equation}
\sqrt{L_n}\left( \frac{ S_{L_n}(1)}{L_n} - \mu(1)\right) \Longrightarrow_n \sigma_0 Z
\label{eq:CLThm+BernoulliSum}
\end{equation}
whence
\begin{equation}
\left ( \frac{ \Gamma(1)}{\Gamma(0)}  \right )^{  \sqrt{L_n}  \left ( \frac{S_{L_n}(1)}{L_n}  - \mu(1) \right )}
 \Longrightarrow_n 
 \left ( \frac{\Gamma(1)}{\Gamma(0)} \right )^{\sigma_0 Z }
\label{eq:CLThm+BernoulliSumExponential}
\end{equation}
by the Continuous Mapping Theorem for weak convergence, see e.g.,  \cite[p. 316]{GrimmettStirzaker}.
Collecting (\ref{eq:LimitA}), (\ref{eq:ConvergenceRatio+B}) and (\ref{eq:CLThm+BernoulliSumExponential}) we readily conclude 
\[
(D_{n,L_n}(1) )^{\frac{1}{\sqrt{L_n}}} e^{-\sqrt{L_n}\left(\frac{1}{\rho_n} + \ln \Gamma(1)^{\mu(1)}\Gamma(0)^{\mu(0)} \right ) } 
 \Longrightarrow_n 
 \left ( \frac{\Gamma(1)}{\Gamma(0)} \right )^{\sigma_0 Z } 
\]
by standard facts concerning the modes of convergence for rvs.
Noting that $ \left ( \frac{\Gamma(1)}{\Gamma(0)} \right )^{\sigma_0 Z }  =e^{ \sigma Z }$, 
this last convergence coincides with (\ref{eq:LOGNORMAL}) and this concludes the proof of Theorem \ref{thm:LOGNORMAL}.
\myendpf

\section{Towards a Berry-Esseen type estimate}
\label{sec:DECOMPOSITION}

Fix $ x \geq 0$ and $n=2,3, \ldots $.
In order to take advantage of the Berry-Esseen complement to the classical Central Limit Theorem, we return to
the proof of Theorem \ref{thm:LOGNORMAL}: To simplify the notation we write
\begin{equation}
R \equiv \frac{ \Gamma(1) }{ \Gamma(0) },
\label{eq:R}
\end{equation}
so that $\sigma = \sigma_0 \cdot \ln R $.
We note from (\ref{eq:RELATION+1}) that
\begin{eqnarray}
\lefteqn{ 
\bP{ (D_{n,L_n}(1) )^{\frac{1}{\sqrt{L_n}}}e^{-\sqrt{L_n}\left(\frac{1}{\rho_n} + \ln \Gamma(1)^{\mu(1)}\Gamma(0)^{\mu(0)} \right ) } \leq x }
} & &
\nonumber \\
&=&
\bP{ 
 \left ( \frac{ D_{n,L_n}(1)}{ \bE{ D_{n,L_n} (1) | S_{L_n}(1) } }  \right )^{\frac{1}{\sqrt{L_n}}}
\cdot
R^{  \sqrt{L_n}  \left ( \frac{S_{L_n}(1)}{L_n}  - \mu(1) \right )}
\leq y_n(x)
} 
\end{eqnarray}
where we have set
\[
y_n (x) \equiv  \left ( \frac{n}{n-1} \right )^{\frac{1}{\sqrt{L_n}}}  \cdot x.
\]

Next we proceed by decomposing $\Delta_n(x)$ given at (\ref{eq:DELTA}) as a sum of three terms, namely
\begin{equation}
R_n (x) = A_n (x) + B_n(x) + C_n(x)
\label{eq:BasicDecomposition}
\end{equation}
with the quantities $A_n(x)$, $B_n(x)$ and $C_n(x)$ given by
\begin{eqnarray}
A_n(x) 
&\equiv & \bP{ 
 \left ( \frac{ D_{n,L_n}(1)}{ \bE{ D_{n,L_n} (1) | S_{L_n}(1) } }  \right )^{\frac{1}{\sqrt{L_n}}}
\cdot
R^{  \sqrt{L_n}  \left ( \frac{S_{L_n}(1)}{L_n}  - \mu(1) \right )}
\leq y_n(x)
} 
\nonumber  \\
& &
~- \bP{ 
R^{  \sqrt{L_n}  \left ( \frac{S_{L_n}(1)}{L_n}  - \mu(1) \right )}
\leq y_n(x)
} ,
\label{eq:A_n(x)}
\end{eqnarray}
\begin{eqnarray}
B_n(x) 
\equiv 
\bP{ R^{  \sqrt{L_n}  \left ( \frac{S_{L_n}(1)}{L_n}  - \mu(1) \right )} \leq y_n(x) } 
 - \bP{  R^{\sigma_0 Z}  \leq y_n(x) } ,
\label{eq:B_n(x)}
\end{eqnarray}
and
\begin{eqnarray}
C_n(x) 
\equiv 
\bP{ R^{\sigma_0 Z} \leq y_n(x) }  - \bP{   R ^{\sigma_0 Z} \leq x } ,
\label{eq:C_n(x)}
\end{eqnarray}
respectively. These terms will be analyzed in turn in the next sections.

\section{Auxiliary results}
\label{sec:AuxiliaryResults}

We begin with easy estimates on various probabilities involving the log-normal distribution.

\begin{lemma}
{\sl
We have the bound
\begin{equation}
\bP{ u < e^{\sigma Z} \leq v }
\leq 
\frac{1}{\sqrt{2\pi \sigma^2}}   \ln \left ( \frac{v}{u} \right ),
\quad 0 < u < v.
\label{eq:BOUND1}
\end{equation}
}
\label{lem:BOUND1}
\end{lemma}

\myproof
Fix $0 < u < v $.
The rvs $\sigma Z$ and $|\sigma|Z$ being equidistributed, we obtain
\begin{eqnarray}
\bP{ u < e^{\sigma Z} \leq v }
=
\bP{ \frac{ \ln u}{|\sigma|}  < Z \leq \frac{ \ln v }{|\sigma|} }
= \Phi \left ( \frac{ \ln v }{|\sigma|} \right ) - \Phi \left ( \frac{ \ln u}{|\sigma|} \right ) .
\nonumber
\end{eqnarray}
The bound (\ref{eq:BOUND1}) follows upon noting that for each $\Delta > 0$, we have
\[
\Phi(t+\Delta ) - \Phi(t)
= \int_t^{t+\Delta} \phi(s)ds
\leq \frac{\Delta}{\sqrt{2\pi}},
\quad t \in \mathbb{R}
\]
by a crude bounding argument (since $\phi(s) \leq \phi(0)$ for all $s$ in $\mathbb{R}$).
\myendpf

Next we present an easy implication  of the classical Berry-Esseen bound.

\begin{lemma}
{\sl Assume $\Gamma(0) \neq \Gamma(1)$.
There exists a constant $C^\star > 0$ such that for every $L=1,2, \ldots $, we have
\begin{eqnarray}
\left |
 \bP{ R^{ \sqrt{L} \left ( \frac{S_L(1)}{L} - \mu(1) \right )}  \leq v } - \bP{ e^{\sigma Z} \leq v }
 \right |
\leq
\frac{C^\star}{\sqrt{L}} \cdot \frac{ \mu(1)^2 + \mu(0)^2 }{ \sqrt{\mu(1)\mu(0)}},
\quad v > 0 .
\label{eq:BOUND2}
\end{eqnarray}
}
\label{lem:BOUND2}
\end{lemma}
It is a simple matter to check from (\ref{eq:BOUND2}) that the following estimate
\begin{eqnarray}
\left |
 \bP{ u < R^{  \sqrt{L} \left ( \frac{S_L(1)}{L} - \mu(1) \right )}  \leq v } - \bP{ u < e^{\sigma Z} \leq v }
 \right |
 \leq
\frac{2C^\star}{\sqrt{L}} \cdot \frac{ \mu(1)^2 + \mu(0)^2 }{ \sqrt{\mu(1)\mu(0)}},
\quad 0< u < v 
\label{eq:BOUND2b}
\end{eqnarray}
also holds for each $L=1,2, \ldots $.

\myproof
The i.i.d. rvs $\{ A_\ell - \mu(1) , \ \ell =1,2, \ldots \}$ have zero mean, variance $\sigma_0^2$ and absolute centered third moment given by
\begin{eqnarray}
\bE{ \left | A_\ell - \mu(1)  \right |^3 }
= \mu(1)\mu(0) \left ( \mu(0)^2 + \mu(1)^2 \right ),
\quad \ell =1,2, \ldots 
\end{eqnarray}
Fix $L=1,2, \ldots $. The Berry-Esseen Theorem  \cite[Section 6, p. 342]{ShiryayevBook} applied to 
the i.i.d. rvs $\{ A_\ell - \mu(1) , \ \ell =1,2, \ldots \}$ yields
\begin{eqnarray}
\sup_{ t \in \mathbb{R} }
\left |
\bP{  \frac{ S_{L} - L \mu(1) }{ \sigma_0 \sqrt{L} } \leq  t }
-
\bP{  Z \leq  t }
\right |
&\leq &
\frac{C^\star}{ \sqrt{L}} \cdot \frac{ \mu(1)\mu(0) \left ( \mu(1)^2 + \mu(0)^2  \right ) }{ \left ( \sqrt{ \mu(1) \mu(0) } \right )^3 }
\nonumber \\
&=&
\frac{C^\star}{\sqrt{L}} \cdot \frac{ \mu(1)^2 + \mu(0)^2 }{ \sqrt{\mu(1)\mu(0)}}
\label{eq:BERRYESSEEN}
\end{eqnarray}
for some universal constant $C^\star > 0$ (independent of $L$).

Next, pick $v > 0$.
Assume first that $\Gamma (0) < \Gamma(1)$ so that $R > 1$ and $\sigma = \sigma_0 \ln R > 0$.
For each $v > 0$ we have
\begin{eqnarray}
\lefteqn{ \bP{ R^{  \sqrt{L} \left ( \frac{S_L(1)}{L} - \mu(1) \right )}  \leq v } - \bP{ e^{\sigma Z} \leq v } } & &
\nonumber \\
&=&
\bP{ \sqrt{L} \left ( \frac{S_L(1)}{L} - \mu(1) \right ) \cdot \ln R   \leq \ln v } - \bP{ \sigma Z  \leq \ln v }
\nonumber \\
&=&
\bP{  \frac{S_L(1) - L \mu(1) }{\sigma_0 \sqrt{L} }   \leq \frac{ \ln v }{ \sigma } } - \bP{  Z  \leq \frac{ \ln v }{ \sigma} },
\end{eqnarray}
and we obtain (\ref{eq:BOUND2}) by the Berry-Esseen bound (\ref{eq:BERRYESSEEN}).
The case $\Gamma (1) < \Gamma (0)$ is handled in a similar way; details are left to the interested reader.
\myendpf

Combining Lemma \ref{lem:BOUND1} and Lemma \ref{lem:BOUND2} (in the form (\ref{eq:BOUND2b})) readily yields the following bound.

\begin{lemma}
{\sl Assume $\Gamma(0) \neq \Gamma(1)$.
For every $L=1,2, \ldots $, we have
\begin{eqnarray}
\bP{ u < R^{ \sqrt{L} \left ( \frac{S_L(1)}{L} - \mu(1) \right )}  \leq v } 
\leq
\frac{2C^\star}{\sqrt{L}} \cdot \frac{ \mu(1)^2 + \mu(0)^2 }{ \sqrt{\mu(1)\mu(0)}}
+ 
\frac{1}{\sqrt{2\pi \sigma^2}}   \ln \left ( \frac{v}{u} \right ),
\quad 0 < u < v
\label{eq:BOUND3}
\end{eqnarray}
where $C^\star$ is the positive constant appearing in Lemma \ref{lem:BOUND2}.
}
\label{lem:BOUND3}
\end{lemma}

\section{A proof of Theorem \ref{thm:BerryEsseenLogNormal}}
\label{sec:ProofTheoremBerryEsseenLogNormal}

Consider a $\rho$-admissible scaling $\myvec{L} : \mathbb{N}_0 \rightarrow  \mathbb{N}_0$ for some $\rho > 0$.
Fix $n=2,, \ldots $ and $ x \geq 0$.
Our point of departure is the decomposition (\ref{eq:BasicDecomposition})  of $\Delta_n(x)$ with $A_n(x)$, $B_n(x)$ and $C_n(x)$
given by (\ref{eq:A_n(x)}), (\ref{eq:B_n(x)}) and (\ref{eq:C_n(x)}), respectively. 
It is plain that
\begin{equation}
|\Delta_n (x)| \leq  |A_n (x)| + |B_n(x)| + |C_n(x)|.
\label{eq:BasicDecompositionb}
\end{equation}

Applying Lemma \ref{lem:BOUND1} with $u= x$ and $v=y_n(x)$ gives
\begin{eqnarray}
0 < C_n(x)
= \bP{ x < e^{\sigma Z } \leq y_n(x) }
&\leq& 
\frac{1}{\sqrt{2\pi \sigma^2}}   \ln \left ( \frac{y_n(x)}{x} \right )
\nonumber \\
&=&  \frac{1}{\sqrt{2\pi \sigma^2 L_n}}  \cdot \ln \left ( \frac{n}{n-1} \right ).
\label{eq:WWW+c}
\end{eqnarray}
Next, we conclude from Lemma \ref{lem:BOUND2} (with $v=y_n(x)$) that
\begin{eqnarray}
\left | B_n(x)  \right |
=
\left | \bP{ R^{\sqrt{L_n} \left ( \frac{S_{L_n}(1)}{L_n} - \mu(1) \right )} \leq y_n(x) } - \bP{ e^{\sigma Z} \leq y_n (x) } \right |
\leq 
\frac{C^\star}{\sqrt{L_n}} \cdot \frac{ \mu(1)^2 + \mu(0)^2 }{ \sqrt{\mu(1)\mu(0)}}.
\label{eq:WWW+b}
\end{eqnarray}

In Section \ref{sec:ProofLemmaBoundA_n(x)} we establish  the following bound.
\begin{lemma}
{\sl Consider a $\rho$-admissible scaling $\myvec{L} : \mathbb{N}_0 \rightarrow  \mathbb{N}_0$ for some $\rho > 0$.
For each $n=2,3, \ldots $, with $\delta$ in $(0,1)$, we have the bound
\begin{eqnarray}
|A_n(x)|
&\leq& 
\bP{ \Bigg  | \frac{D_{n,L_n}(1) }{\bE{D_{n,L_n}(1) |S_{L_n}(1)}}  -1 \Bigg | > \delta }  
\label{eq:BoundForA_n(x)} \\
& &
~+ \bP{  \left ( 1+\delta \right )^{-\frac{1}{\sqrt{L_n}}} \cdot y_n(x) < R^{\sqrt{L_n} \left ( \frac{S_{L_n}(1)}{L_n} - \mu(1) \right )} 
\leq \left ( 1-\delta \right )^{-\frac{1}{\sqrt{L_n}}} \cdot y_n(x) }  .
\nonumber 
\end{eqnarray}
}
\label{lem:BoundForA_n(x)}
\end{lemma}

Lemma \ref{lem:BOUND3} (with $u = \left ( 1+\delta \right )^{-\frac{1}{\sqrt{L_n}}}  \cdot y_n(x)$ 
and $v = \left ( 1-\delta \right )^{-\frac{1}{\sqrt{L_n}}} \cdot y_n(x)$ where $0 < \delta < 1$)
already gives
\begin{eqnarray}
& &
\bP{ \left ( 1+\delta \right )^{-\frac{1}{\sqrt{L_n}}} \cdot y_n(x)   < R^{ \sqrt{L_n} \left ( \frac{S_{L_n} (1)}{L_n} - \mu(1) \right )} 
\leq \left ( 1-\delta \right )^{-\frac{1}{\sqrt{L_n}}} \cdot y_n(x)  } 
\nonumber \\
&\leq&
\frac{1}{\sqrt{2\pi \sigma^2 L_n}}   \ln \left ( \frac{1+ \delta}{1-\delta} \right ) 
+
\frac{2C^\star}{\sqrt{L_n}} \cdot \frac{ \mu(1)^2 + \mu(0)^2 }{ \sqrt{\mu(1)\mu(0)}} .
\nonumber
\end{eqnarray}
On the other hand, 
when used along the scaling $\rho$-admissible scaling $\myvec{L} : \mathbb{N}_0 \rightarrow  \mathbb{N}_0$,
Lemma \ref{lem:ConvergenceRatio+Rate} gives 
\[
\bP{ \Bigg | \frac{ D_{n,L_n}(1) }{ \bE{ D_{n,L_n}(1) | S_{L_n} (1) }  } - 1 \Bigg |  ~> \delta }
\leq
4 e^{ -2 L_n \eta^2 }
+
2 e^{ - \Psi (\delta) \cdot (n-1) \left ( \Gamma(1)^{\mu(1) + \eta }  \Gamma(0)^{1 - \mu(1) + \eta} \right )^{L_n} }
\]
with $\eta$ in $(0, \mu(1))$. Combining the last two inequalities we conclude from Lemma \ref{lem:BoundForA_n(x)} that
\begin{eqnarray}
| A_n(x) | 
&\leq &
\frac{1}{\sqrt{2\pi \sigma^2 L_n}}  \cdot \ln \left ( \frac{1+\delta}{1-\delta} \right )
+
\frac{2C^\star}{\sqrt{L_n}} \cdot \frac{ \mu(1)^2 + \mu(0)^2 }{ \sqrt{\mu(1)\mu(0)}}
\nonumber \\
& &
~ + 4 e^{ -2 L_n \eta^2 }
+
2 e^{ - \Psi (\delta) \cdot (n-1) \left ( \Gamma(1)^{\mu(1) + \eta }  \Gamma(0)^{1 - \mu(1) + \eta} \right )^{L_n} }.
\label{eq:WWW+a}
\end{eqnarray}
The proof of Theorem \ref{thm:BerryEsseenLogNormal} is completed by applying the bounds 
(\ref{eq:WWW+c}), (\ref{eq:WWW+b}) and (\ref{eq:WWW+a}) in the decomposition (\ref{eq:BasicDecompositionb}),
and collecting like terms.
\myendpf

\section{A proof of Lemma \ref{lem:BoundForA_n(x)}}
\label{sec:ProofLemmaBoundA_n(x)}

For each $n=2,3, \ldots $, write
\[
R_n \equiv 
\frac{ D_{n,L_n}(1)}{ \bE{ D_{n,L_n} (1) | S_{L_n}(1) } }  
\quad \mbox{and} \quad 
{\rm CLT}_n
\equiv  \sqrt{L_n}  \left ( \frac{S_{L_n}(1)}{L_n}  - \mu(1) \right )
\]
with $R$ given by (\ref{eq:R}).
Fix $x \geq 0$. We note that
\begin{eqnarray}
A_n(x) 
&\equiv & 
\bP{   \left ( R_n  \right )^{\frac{1}{\sqrt{L_n}}} \cdot R^{{\rm CLT}_n}  \leq y_n(x) } 
-
\bP{   R^{{\rm CLT}_n}  \leq y_n(x) } 
\nonumber  \\
&=& 
\bE{ \1{  \left ( R_n  \right )^{\frac{1}{\sqrt{L_n}}} \cdot R^{{\rm CLT}_n}  \leq y_n(x) }  
-
\1{ R^{{\rm CLT}_n}  \leq y_n(x) }
}.
\label{eq:AAA}
\end{eqnarray}

We start with the obvious decompositions
\begin{eqnarray}
\lefteqn{
\1{  \left ( R_n  \right )^{\frac{1}{\sqrt{L_n}}} \cdot R^{{\rm CLT}_n}  \leq y_n(x) }  
} & &
\nonumber \\
&=&
\1{ R_n \leq 1,  \left ( R_n  \right )^{\frac{1}{\sqrt{L_n}}} \cdot R^{{\rm CLT}_n}  \leq y_n(x) }  
+
\1{ 1 < R_n ,  \left ( R_n  \right )^{\frac{1}{\sqrt{L_n}}} \cdot R^{{\rm CLT}_n}  \leq y_n(x) }  
\label{eq:AAA+1}
\end{eqnarray}
and
\begin{eqnarray}
\1{  R^{{\rm CLT}_n}  \leq y_n(x) }  
=
\1{ R_n \leq 1,  R^{{\rm CLT}_n}  \leq y_n(x) }  
+
\1{ 1 < R_n ,   R^{{\rm CLT}_n}  \leq y_n(x) }  .
\label{eq:AAA+2}
\end{eqnarray}
Substracting (\ref{eq:AAA+2}) from (\ref{eq:AAA+1}) and collecting like terms we get
\begin{eqnarray}
\lefteqn{
 \1{  \left ( R_n  \right )^{\frac{1}{\sqrt{L_n}}} \cdot R^{{\rm CLT}_n}  \leq y_n(x) }  
-
\1{ R^{{\rm CLT}_n}  \leq y_n(x) }
} & &
\nonumber \\
&=&
\1{ R_n \leq 1 }
\left ( \1{  \left ( R_n  \right )^{\frac{1}{\sqrt{L_n}}} \cdot R^{{\rm CLT}_n}  \leq y_n(x) }  -  \1{R^{{\rm CLT}_n}  \leq y_n(x) }  \right )
\nonumber \\
& &
~+
\1{ 1 < R_n }
\left ( \1{  \left ( R_n  \right )^{\frac{1}{\sqrt{L_n}}} \cdot R^{{\rm CLT}_n}  \leq y_n(x) }  -  \1{R^{{\rm CLT}_n}  \leq y_n(x) }  \right ).
\label{eq:AAA+3}
\end{eqnarray}

The first terms in (\ref{eq:AAA+3})  can be rewritten as
\begin{eqnarray}
\lefteqn{
\1{ R_n \leq 1 }
\left ( \1{  \left ( R_n  \right )^{\frac{1}{\sqrt{L_n}}} \cdot R^{{\rm CLT}_n}  \leq y_n(x) }  -  \1{R^{{\rm CLT}_n}  \leq y_n(x) }  \right )
} & & 
\nonumber \\
&=&
\1{ R_n \leq 1 }
\1{  \left ( R_n  \right )^{\frac{1}{\sqrt{L_n}}} \cdot R^{{\rm CLT}_n}  \leq y_n(x), y_n(x) <  R^{{\rm CLT}_n} }
\nonumber \\
& & +
\1{ R_n \leq 1 }
\left (
\1{  \left ( R_n  \right )^{\frac{1}{\sqrt{L_n}}} \cdot R^{{\rm CLT}_n}  \leq y_n(x), R^{{\rm CLT}_n} \leq y_n(x) }  
-  \1{ R^{{\rm CLT}_n}  \leq y_n(x) }  \right )
\nonumber \\
&=&
\1{ R_n \leq 1 }
\1{  \left ( R_n  \right )^{\frac{1}{\sqrt{L_n}}} \cdot R^{{\rm CLT}_n}  \leq y_n(x) <  R^{{\rm CLT}_n} }
\nonumber \\
& & -
\1{ R_n \leq 1 }
\1{  R^{{\rm CLT}_n} \leq y_n(x) < \left ( R_n  \right )^{\frac{1}{\sqrt{L_n}}} \cdot R^{{\rm CLT}_n} }  
\nonumber \\
&=&
\1{ R_n \leq 1 }
\1{  \left ( R_n  \right )^{\frac{1}{\sqrt{L_n}}} \cdot R^{{\rm CLT}_n}  \leq y_n(x) <  R^{{\rm CLT}_n} }
\label{eq:AAA+4}
\end{eqnarray}
while the second term in (\ref{eq:AAA+3}) becomes
\begin{eqnarray}
\lefteqn{
\1{ 1 < R_n }
\left ( \1{  \left ( R_n  \right )^{\frac{1}{\sqrt{L_n}}} \cdot R^{{\rm CLT}_n}  \leq y_n(x) }  -  \1{R^{{\rm CLT}_n}  \leq y_n(x) }  \right )
} & &
\nonumber \\
&=&
\1{ 1 < R_n }
 \1{  \left ( R_n  \right )^{\frac{1}{\sqrt{L_n}}} \cdot R^{{\rm CLT}_n}  \leq y_n(x), y_n(x) < R^{{\rm CLT}_n}   }  
\nonumber \\
& &
+
\1{ 1 < R_n }
\left ( \1{  \left ( R_n  \right )^{\frac{1}{\sqrt{L_n}}} \cdot R^{{\rm CLT}_n}  \leq y_n(x),  R^{{\rm CLT}_n} \leq y_n(x)  }  -  \1{R^{{\rm CLT}_n}  \leq y_n(x) }  \right )
\nonumber \\
&=&
\1{ 1 < R_n }
 \1{  \left ( R_n  \right )^{\frac{1}{\sqrt{L_n}}} \cdot R^{{\rm CLT}_n}  \leq y_n(x) < R^{{\rm CLT}_n}   }  
\nonumber \\
& &
-
\1{ 1 < R_n }
 \1{  R^{{\rm CLT}_n} \leq y_n(x)  < \left ( R_n  \right )^{\frac{1}{\sqrt{L_n}}} \cdot R^{{\rm CLT}_n}  }  
 \nonumber \\
&=&
-
\1{ 1 < R_n }
 \1{  R^{{\rm CLT}_n} \leq y_n(x)  < \left ( R_n  \right )^{\frac{1}{\sqrt{L_n}}} \cdot R^{{\rm CLT}_n}  }  .
 \label{eq:AAA+5}
\end{eqnarray}

Inserting (\ref{eq:AAA+4}) and (\ref{eq:AAA+5}) back into (\ref{eq:AAA+3}), and take expectations.
With the help of  (\ref{eq:AAA}) we then conclude that
\begin{eqnarray}
A_n(x) 
&=&
\bE{
\1{ R_n < 1 }
\1{  \left ( R_n  \right )^{\frac{1}{\sqrt{L_n}}} \cdot R^{{\rm CLT}_n}  \leq y_n(x) <  R^{{\rm CLT}_n} }
}
\nonumber \\
& & - \bE{ \1{ 1 < R_n }
 \1{  R^{{\rm CLT}_n} \leq y_n(x)  < \left ( R_n  \right )^{\frac{1}{\sqrt{L_n}}} \cdot R^{{\rm CLT}_n}  }  
 }
 \nonumber \\
 &=&
\bP{ R_n < 1,  \left ( R_n  \right )^{\frac{1}{\sqrt{L_n}}} \cdot R^{{\rm CLT}_n}  \leq y_n(x) <  R^{{\rm CLT}_n} }
\nonumber \\
& & - \bP{ 1 < R_n ,  R^{{\rm CLT}_n} \leq y_n(x)  < \left ( R_n  \right )^{\frac{1}{\sqrt{L_n}}} \cdot R^{{\rm CLT}_n}  }  .
 \label{eq:AAA+6}
 \end{eqnarray}

Pick $\delta $ arbitrary in $(0,1)$.
It follows that
\begin{eqnarray}
|A_n(x)|
&\leq& 
\bP{ R_n < 1-\delta,  \left ( R_n  \right )^{\frac{1}{\sqrt{L_n}}} \cdot R^{{\rm CLT}_n}  \leq y_n(x) <  R^{{\rm CLT}_n} }
\nonumber \\
& & 
~+ \bP{ 1 - \delta \leq R_n < 1,  \left ( R_n  \right )^{\frac{1}{\sqrt{L_n}}} \cdot R^{{\rm CLT}_n}  \leq y_n(x) <  R^{{\rm CLT}_n} }
\nonumber \\
& &
~+ \bP{ 1 < R_n \leq 1+\delta ,  R^{{\rm CLT}_n} \leq y_n(x)  < \left ( R_n  \right )^{\frac{1}{\sqrt{L_n}}} \cdot R^{{\rm CLT}_n}  }  
\nonumber \\
& &
~+ \bP{ 1+\delta  < R_n ,  R^{{\rm CLT}_n} \leq y_n(x)  < \left ( R_n  \right )^{\frac{1}{\sqrt{L_n}}} \cdot R^{{\rm CLT}_n}  }  
\nonumber \\
&\leq& \bP{ | R_n -1 | > \delta } 
+ \bP{ 1 - \delta \leq R_n < 1,  \left ( R_n  \right )^{\frac{1}{\sqrt{L_n}}} \cdot R^{{\rm CLT}_n}  \leq y_n(x) <  R^{{\rm CLT}_n} }
\nonumber \\
& &
~+ \bP{ 1 < R_n \leq 1+\delta ,  R^{{\rm CLT}_n} \leq y_n(x)  < \left ( R_n  \right )^{\frac{1}{\sqrt{L_n}}} \cdot R^{{\rm CLT}_n}  }  .
 \label{eq:AAA+7}
\end{eqnarray}
It is now straightforward to see that
\begin{eqnarray}
& & \bP{ 1 - \delta \leq R_n < 1,  \left ( R_n  \right )^{\frac{1}{\sqrt{L_n}}} \cdot R^{{\rm CLT}_n}  \leq y_n(x) <  R^{{\rm CLT}_n} }
\nonumber \\
& \leq&
\bP{ 1 - \delta \leq R_n < 1,  \left ( 1-\delta \right )^{\frac{1}{\sqrt{L_n}}} \cdot R^{{\rm CLT}_n}  \leq y_n(x) <  R^{{\rm CLT}_n} }
\nonumber \\
&=&
\bP{ 1 - \delta \leq R_n < 1,  y_n(x) < R^{{\rm CLT}_n}  \leq \left ( 1-\delta \right )^{-\frac{1}{\sqrt{L_n}}} \cdot y_n(x)  }
\nonumber \\
& \leq&
\bP{  y_n(x) < R^{{\rm CLT}_n}  \leq \left ( 1-\delta \right )^{-\frac{1}{\sqrt{L_n}}} \cdot y_n(x) }
 \label{eq:AAA+8}
\end{eqnarray}
and
\begin{eqnarray}
& & \bP{ 1 < R_n \leq 1+\delta ,  R^{{\rm CLT}_n} \leq y_n(x)  < \left ( R_n  \right )^{\frac{1}{\sqrt{L_n}}} \cdot R^{{\rm CLT}_n}  }  
\nonumber \\
& \leq&
\bP{ 1 < R_n \leq 1+\delta ,  R^{{\rm CLT}_n} \leq y_n(x)  < \left ( 1+\delta \right )^{\frac{1}{\sqrt{L_n}}} \cdot R^{{\rm CLT}_n}  }  
\nonumber \\
&=&
\bP{ 1 < R_n \leq 1+\delta ,   \left ( 1+\delta \right )^{-\frac{1}{\sqrt{L_n}}} \cdot y_n(x)  < R^{{\rm CLT}_n} \leq y_n(x) }  
\nonumber \\
&\leq&
\bP{  \left ( 1+\delta \right )^{-\frac{1}{\sqrt{L_n}}} \cdot y_n(x)  < R^{{\rm CLT}_n} \leq y_n(x) }  
 \label{eq:AAA+9}
\end{eqnarray}

Applying the bounds  (\ref{eq:AAA+8}) and (\ref{eq:AAA+9})  to (\ref{eq:AAA+7}) readily leads to the inequality
(\ref{eq:BoundForA_n(x)}), and this completes the proof of Lemma \ref{lem:BoundForA_n(x)}.
\myendpf

\section*{Acknowledgment}
This work was supported by NSF Grant CCF-1217997.
The paper was completed during the academic year 2014-2015  while A.M. Makowski 
was a Visiting Professor with the Department of Statistics of the Hebrew University of Jerusalem 
with the support of a fellowship from the Lady Davis Trust.

\bibliographystyle{IEEE}

\end{document}